\documentclass[pdftex,twocolumn]{aastex631}
\usepackage{amssymb,amsmath}
\usepackage{color}

\graphicspath{{./}{figures/}}

\def\degs{\ifmmode ^{\circ}\else$^{\circ}$\fi}
\def\amin{\ifmmode ^{\prime}\else$^{\prime}$\fi}
\def\asec{\ifmmode ^{\prime\prime}\else$^{\prime\prime}$\fi}
\def\h{$^{\rm h}$}
\def\m{$^{\rm m}$}
\def\s{$^{\rm s}$}

\usepackage[normalem]{ulem}  % \sout{old text} for strikeout
\usepackage{color,soul}

\usepackage{hyperref}
\usepackage{url}
\usepackage{natbib}

\begin{document}

\title{
%HUBBLE SPACE TELESCOPE OBSERVATIONS OF THE OLD PULSAR B0950+08
Thermal and nonthermal emission in the optical-UV spectrum of PSR B0950+08\footnote{Based on observations made with 
the NASA/ESA {\sl Hubble Space Telescope}, obtained at the Space Telescope Science Institute, which is operated by the Association 
of Universities for Research in Astronomy, Inc., under NASA contract NAS 5-26555. These observations are associated with program \#16064.} 
}

%% Note that the corresponding author command and emails has to come
%% before everything else. Also place all the emails in the \email
%% command instead of using multiple \email calls.
\correspondingauthor{George G.\ Pavlov}
\email{ggp1@psu.edu}

\author{Vadim Abramkin}
%\author{Second Author}
%\altaffiliation{Editor-in-Chief}
\affiliation{Ioffe Institute, Politekhnicheskaya 26, St.\ Petersburg, 195251, Russia}
\author{George G.\ Pavlov}
\affiliation{Pennsylvania State University, Department of Astronomy \& Astrophysics, 525 Davey Lab., University Park, PA 16802, USA}
\author{Yuriy Shibanov}
%\altaffiliation{RNAAS Editor}
\affiliation{Ioffe Istitute, Politekhnicheskaya 26, St.\ Petersburg, 195251, Russia}

%% The \author command can take an optional ORCID.
%\author[0000-0002-0786-7307]{Third Author  }
\author{Oleg Kargaltsev}
%\affiliation{INAF -- Istituto di Astrofisica Spaziale e Fisica Cosmica Milano, via E. Bassini 15, I-20133, Milano, Italy}
\affiliation{The George Washington University, Department of Physics, 725 21st Street NW, Washington, DC 20052, USA
}
%Janusz Gil Institute of Astronomy, University of Zielona G\'{o}ra, ul Szafrana 2, 65-265, Zielona G\'{o}ra, Poland}

%\author{Forth Author}
%\affiliation{Pennsylvania State University, Department of Astronomy \& Astrophysics, 525 Davey Lab., University Park, PA 16802, USA}

%% Note that RNAAS manuscripts DO NOT have abstracts.

\begin{abstract}
Based on recent Hubble Space Telescope (HST) observations in far-UV 
%bands 
and groundbased observations
in optical bands, Pavlov and colleagues have revealed a thermal component in the spectrum of the %nearby, 
old pulsar B0950+08 
%($d\simeq 262$ pc, 
(spin-down age 17.5 Myr)
and estimated a neutron star (NS) surface temperature of $(1$--$3)\times 10^5$ K. Our new HST observations in the optical have allowed us to resolve the pulsar from a close-by galaxy and measure  the
optical fluxes more accurately. Using the newly measured fluxes and a new calibration of the HST's far-UV detector, we fit the optical-UV pulsar's spectrum 
%$f_\nu$ 
with a 
model that consists of a nonthermal %(presumably magnetospheric) 
power-law %component 
($f_\nu\propto \nu^\alpha$) and a thermal blackbody 
%spectral 
components.
%emitted from the %neutron star 
%pulsar's magnetosprere and the NS surface, respectively. 
We obtained the spectral slope $\alpha=-0.3\pm 0.3$, considerably flatter than found from groundbased observations, and the best-fit %blackbody 
temperature in the range of %$(5$--$15)\times 10^4$ K [???]
%\va{
$(6$--$12)\times 10^4$ K (as seen by a distant observer), depending on interstellar extinction and  NS radius.
The temperature is lower than  
reported previously, 
but still much higher than predicted by %neutron star 
NS passive cooling scenarios for such an old pulsar. 
This means that some heating mechanisms  operate in NSs,
%neutron stars, %such as due to 
e.g., caused by interaction %of vortices 
of the faster rotating neutron superfluid with the slower rotating normal matter in the inner  crust of the NS. 
\end{abstract}

\section{Introduction}
\label{introduction}

Born very hot, neutron stars (NSs) lose their thermal energy via neutrino and photon emission. Since the cooling rate depends on the composition and state of matter in the NS interiors, the study of thermal evolution of NS is an important tool for understanding the fundamental properties of matter.
While the thermal evolution of young NSs in the neutrino-dominated cooling era ($\tau \lesssim 1$ Myr) has been 
well investigated 
%(see \citet{YakovlevPethick2004} and \citet{Page2009} for reviews),
(see \citealt{YakovlevPethick2004} and \citealt{Page2009} for reviews),
the thermal evolution of older NSs remains virtually unexplored. If an NS just cools passively, then its surface
temperature is expected to drop very fast in the photon-dominated cooling era ($\tau\gtrsim 1$ Myr), going below $10^4$ K by $\tau\sim 10$ Myr. However, it has long been recognized that various
heating processes may slow (or even reverse) the cooling (e.g.,
\citealt{GonzalezReisenegger2010}, and references therein). 

For instance, interaction of vortex lines of the faster rotating
neutron superfluid with the slower rotating normal matter in the
inner NS crust 
can heat the NS surface 
%of $\sim 10$--100 Myr old pulsars 
up to temperatures of a few times $10^5$ K at an age of $\sim 10$--100 Myr
%, depending on the properties of the NS interiors 
(\citealt{Alpar1984}; \citealt{ShibazakiLamb1989}; \citealt{LarsonLink1999}; \citealt{GonzalezReisenegger2010}). This ``frictional heating'' 
mechanism can be explored in observations of thermal emission from old NSs. An optimal range for such observation is far-UV (FUV) because the thermal flux is higher there at the expected temperatures and because optical spectrum can be dominated by magnetospheric emission if the NS is a rotation-powered pulsar. %Thus, 
Since FUV emission is unobservable from the ground, observational studies of heating mechanisms and thermal evolution of NSs can only be done with the Hubble Space Telescope (HST; \citealt{Pavlov1992}). Since NSs in general, and the thermal evolution of NSs in particular, are not the favorite subjects of HST panels, very few proposals on this topic have been accepted in 30 years of the HST era. Nevertheless, those few accepted and executed programs led to important discoveries, one of which was the observational confirmation of NS heating in two old pulsars -- the millisecond (recycled) pulsar J0437--4715 and the ``classical'' (nonrecycled) pulsar B0950+08. New observations of the latter is the main subject of this paper.

PSR B0950+08 (= J0953+0755; B0950 hereafter) is  a
solitary 
radio and X-ray pulsar, 
%{\bf 
one of the first pulsars discovered %in the radio 
\citep{1968Natur}. 
%}
%with
It has the period $P = 253$ ms, spin-down energy loss rate $\dot{E} = 5.6\times 10^{32}$ erg s$^{-1}$,
 spin-down age $\tau_{\rm sd} = 17.5$ Myr, and surface
magnetic field $B\sim 2.4\times  10^{11}$ G \citep{Manchester2005}.
%B0950 has a 
The parallax distance $d=262\pm 5$ pc and 
 proper motion $\mu_\alpha= -2.08 \pm 0.08$ mas yr$^{-1}$, 
$\mu_\delta = 29.46\pm 0.07$  mas yr$^{-1}$
correspond to
the transverse velocity $V_\perp =36.6\pm 0.7$ km s$^{-1}$ \citep{Brisken2002}. %{\bf It was  among first pulsars discovered in the radio \citep{1968Natur}  where it 
%{\bf It shows  giant radio pulses mainly observed at low frequencies around 100 MHz  \citep[e.g.,][and references therein]{2018kazantsev}.} 

Optical–UV emission from B0950 was discovered by \citet{Pavlov1996} in HST observations with the Faint Object Camera 
%(FOC) 
in a long-pass filter F130LP (pivot wavelength 3438\,\AA).%FWHM 2000\,\AA). 
They measured a mean spectral flux density $\overline{f}_\nu = 51\pm 3$ nJy in this filter. In observations with the Subaru telescope 
\citet{Zharikov2002} measured the flux density of $60\pm 9$ nJy in the B filter. Observations with the VLT/FORS1 telescope gave the B, V, R and I 
flux densities of $60\pm 17$, $45\pm 6$, $91\pm8$ and $78\pm 11$ nJy, respectively \citep{Zharikov2004}. The large scatter in the flux values 
was likely caused by contamination, particularly in the I and R bands, from a red extended object at about $1''$ north of the pulsar. 
Despite this scatter, it is clear that the optical emission from B0950 is nonthermal. Fitting these flux densities with a power-law (PL) model, 
$f_\nu \propto \nu^\alpha$, \citet{Zharikov2004} obtained $\alpha=-0.65\pm0.40$.

%%%%%%%%%%%%%%%%%%%%%% Table 1 $$$$$$$$$$$$$$$$$$$$$$$$$$$$
\begin{deluxetable*}{cccccrcc}[t!]
\tablecaption{{\sl HST} observations of PSR B0950+08  \label{tab:obs}}
\tablecolumns{6}
\tablenum{1}
\tablewidth{0pt}
\tablehead{
\colhead{Visit} &
\colhead{Start time} &
%\tablenotemark{a}} &
\colhead{Instrument} &
\colhead{Filter} & 
\colhead{$\lambda_p$\tablenotemark{a} } &
%\colhead{FWHM\tablenotemark{b} } &
\colhead{$\rm W_{eff}$\tablenotemark{b} } &
\colhead{Dither points} &
\colhead{Total exposure}\\
%\tablenotemark{c}} \\
\colhead{
%(yyyy-mm-dd hh:mm:ss)
} & \colhead{ }
%UTC{\bf ??}} 
&
\colhead{} & \colhead{ } &
\colhead{ (\AA)} & 
\colhead{(\AA)} &
\colhead{} &
\colhead{(s)} 
}
\startdata
1 & 2020-05-15\, 11:07:49 & WFC3/UVIS & F475X & 4936 & 1981 & 4 &  2420 \\
2 & 2020-06-05\, 04:25:00 & ACS/WFC & F775W & 7693 & 1379 & 3 & 3726 \\
3 & 2021-03-06\, 15:23:40 & ACS/WFC & F775W & 7693 & 1379 & 2 & 2434 \\
%2016-08-08 19:44:39 & WFC3/UVIS & F336W & 3355 &  2580 \\
%2016-08-08 21:20:04 & WFC3/UVIS & F438W & 4325 & 2580 \\
%2016-08-08 22:55:30 & WFC3/UVIS & F438W & 4325 & 2580 \\
%2016-08-11 22:32:15 & ACS/SBC & F140LP & 1528 & 2800 \\
\enddata
%\tablecomments{\gp{is the time in UTC system?}}
%\tablenotetext{a}{Start time corresponds to the start of first exposure.}
%{Each line 
%is related to a specific 
%corresponds to one {\sl HST}  orbit, 
%and the start time 
%coincides with the start of  
%the first exposure in the orbit.}
\tablenotetext{a}{Pivot wavelength of the filter.}
%\tablenotetext{c}{Exposure time
%per the {\sl HST} orbit.}
\tablenotetext{b}{ %Full width at half maximum of filter throughput. 
Rectangular width of filter throughput.
%\gp{`Rectangular width' may be more appropriate...}
}
\end{deluxetable*}
%%%%%%%%%%%%%%%%%%%%%%%%%%%%%%%%%%%%%%%%%

In a search for thermal emission from the 
NS surface, \citet{Pavlov2017} observed B0950 
%in the far-UV (FUV) range using the 
with the Solar-Blind Channel (SBC) of the HST Advanced Camera for Surveys (ACS). The observations were carried out with the F125LP and F140LP long-pass filters, with pivot wavelengths 1438 \AA\ and 1528 \AA.
%and FWHM ?? and ??, respectively. 
The corresponding mean flux densities, $\overline{f}_\nu = 109\pm 6$ and $83\pm 14$ nJy, substantially exceeded the extrapolation of the optical PL spectrum into this 
%far-UV (FUV) 
FUV range. Therefore, \citet{Pavlov2017} conclude that the FUV excess is due to the thermal emission. Fitting the optical-FUV flux densities with a two-component PL + blackbody (BB) model, they found a PL slope $\alpha\sim - 1.2$ and a brightness temperature in the range of $(1$--$3)\times 10^5$ K, depending on interstellar extinction and NS radius. These temperatures are much higher than predicted by NS cooling models for such an old pulsar, which means that some heating mechanisms operate in NSs.

This conclusion, however, was based on a spectral fit of optical-FUV flux densities, with optical points suffering from poorly known systematic errors caused by contamination from the nearby red extended object in the VLT observations. Moreover, after the work by \citet{Pavlov2017} had been published, \citet{Avila2019} reported a 20\%--30\% error in 
%the absolute flux 
calibration of the SBC detector, which resulted in overestimation of measured fluxes.
To measure the optical spectrum free of contamination caused the nearby extended source and to measure the brightness temperature more accurately, we observed B0950 with HST in two optical filters. Results of these observations and a new fit of the optical-FUV spectrum of B0950 are presented below.

%%%%%%%%%%%%%%%%%%%%%%%%%%%%%%%%%%%%%
\section{Observations}
\label{sec:observations}

B0950 was observed with HST in 3 visits (Program 
16064; 
PI G.\ Pavlov). 
First 
observation was carried out on 2020 May 15
with the Ultraviolet-Visible channel (UVIS; $162''\times 162''$ field of view, 0\farcs04 pixel scale) of the Wide Field Camera 3 (WFC3; 1 orbit) in the extremely wide blue filter F475X 
%($\lambda_p= 4939?$\,\AA, WFHM = 2200?\,\AA). 
(see Table \ref{tab:obs} for the exposure times and filter characteristics). 
%The %target was observed in the 
%exposure was taken in ACCUM mode, applying a 
A four-point box dither pattern was used to filter out cosmic ray events and bad pixels. The target was placed in UVIS2-C512C-CTE aperture,
%was used to place the pulsar 
close to a readout amplifier, to minimize the CCD charge
transfer efficiency (CTE) losses.

Second and third observations,
%was carried out on 2020 June 5
both with the Wide Field Channel (WFC; $202''\times 202''$ field of view, 0\farcs05 pixel scale) of the Advanced Camera for Surveys (ACS) in the F775W (SDSS $i$) filter,
%. This visit was originally planned for 2 orbits, but 
were carried out on 2020 June 5 (1.5 orbits)
and 2021 March 6 (1 orbit),
%. These observations were carried out in ACCUM mode, using 
The target was placed in WFC1-CTE aperture.
The former observation used 3 points of a four-point box dither pattern, while a two-point dither pattern was used in the latter observation.

The data were downloaded from the Barbara A.\ Mikulski Archive for Space Telescopes (MAST)  
and processed using AstroDrizzle software, ver.\ 3.1.6, from the {\tt DrizzlePac} package. 
We chose $0\farcs025\times 0\farcs025$ for the pixel size in the drizzled images
instead of $0\farcs05\times 0\farcs05$ 
%arcsec$^2$ 
%for the image pixel size 
in the MAST pipeline product,
%we chose
%used final scale of 
%$0\farcs025\times 0\farcs025$ %arcsec$^2$
%/pixel for the final image scale
%in order 
which allowed us to increase angular resolution and %easier handle the contamination 
more reliably separate the pulsar counterpart from a nearby extended object (see Sections \ref{astrom} and \ref{phot}).
%\yus{
%After i
Inspection of  
%individual  
separate subexposures and their comparison with the drizzled images revealed
%processed images, 
a few 
%obvious 
cosmic 
ray events and hot pixels  
that remained unfiltered by the drizzling.
%were found to be unfiltered  in the processed images. 
We 
%thus f
filtered them out   manually.
%\gp{describe the drizzling procedures and parameters used....}

\section{Data analysis and results}

\subsection{Astrometry and images}
\label{astrom}

\begin{figure*}[t]
\begin{minipage}[h]{0.5\linewidth}
\center{\includegraphics[width=1\linewidth]{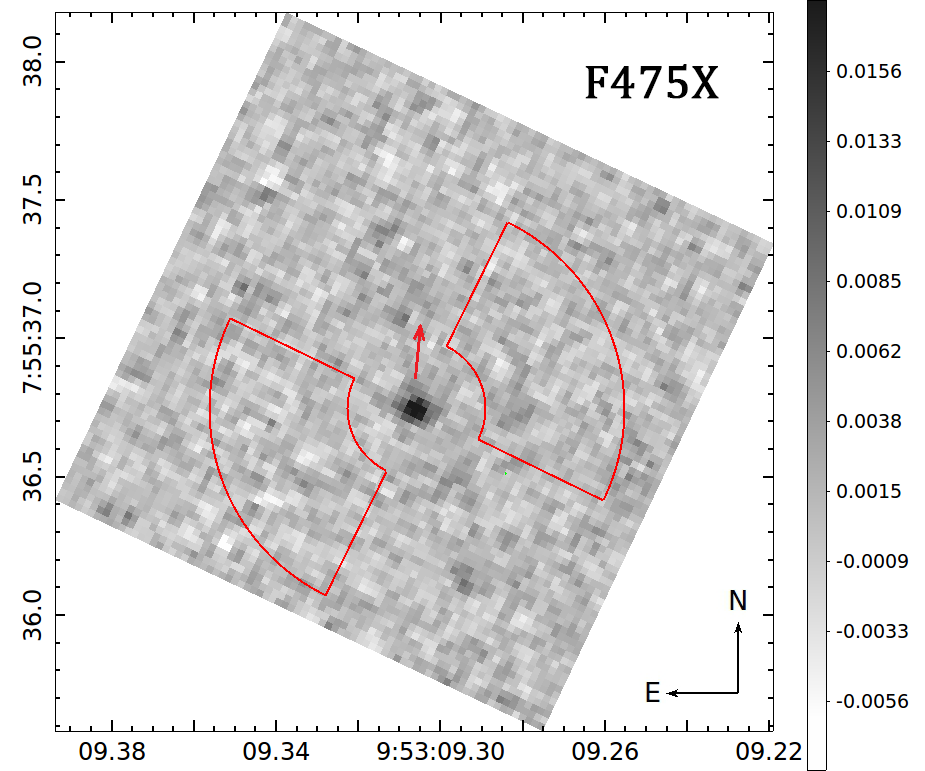}}
\end{minipage}
\hfill
\begin{minipage}[h]{0.5\linewidth}
\center{\includegraphics[width=1\linewidth]{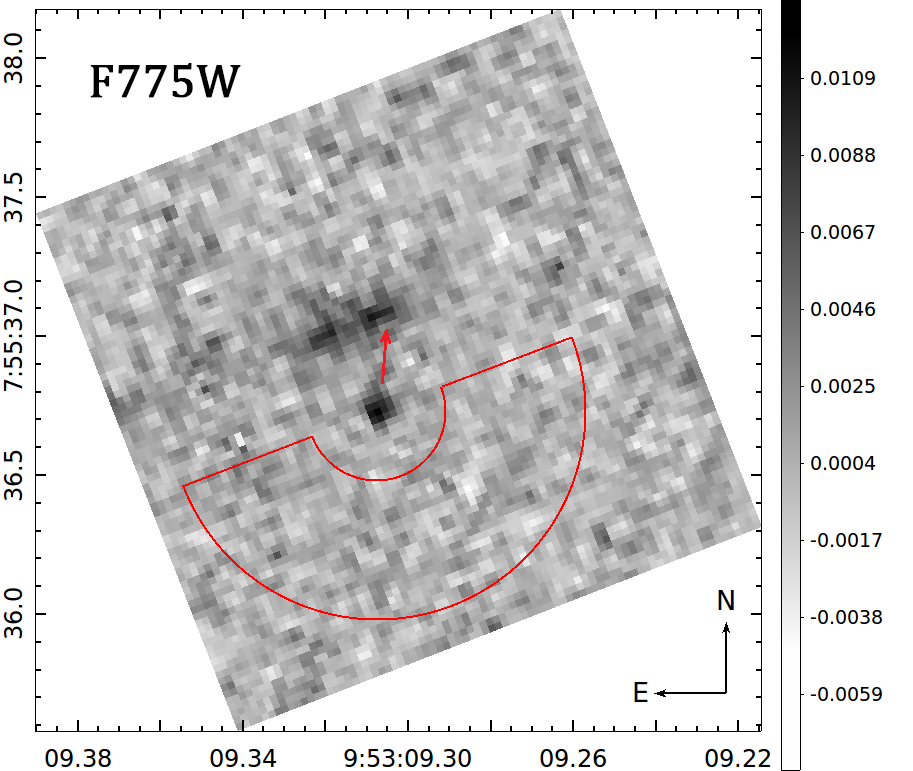}}
\end{minipage}
\caption{
F475X 
%\gp{[looks too dirty now]} 
and F775W images 
%\gp{[would it be possible to put F475X and F775W labels directly onto the images?]}
of $\approx 2\asec \times 2\asec$ pulsar vicinity, with 
%the pixel scale of 
$25\times 25$ mas$^2$ pixels. %\yus{axes titles?}
The pulsar counterpart is the point-like object at the center. The red arrow shows the direction of the pulsar's proper motion; the tip of the arrow shows the position where the pulsar will be in 2031. %\gp{[??]}. 
Red contours show the areas in which the background was estimated.
The vertical bars decode 
the image intensity in counts~s$^{-1}$ per pixel. 
%\gp{what is the pixel size? 25 mas? Was any smoothing applied?} \va{25 mas, no smoothing applied}
%The 0\farcs2 circles are centered on the expected radio pulsar position.
%The 
%red 
%extended background object is seen in the F775W band north-northeast of the pulsar. 
%\gp{Would it be possible/reasonable to show the pulsar proper motion with an arrow? And perhaps show the source and bgd apertures? 
%Would it be possible/reasonable to remove the smudge NNE of the pulsar if we are sure it is not a real source?  Were the two F775W images summed or it is just first image?} \gp{Summed in the absolute coordinates or with pulsar alignment?} 
%\va{with alignment} \gp{then images of field sources become artificially spread?} \va{yes}
%\gp{[a sentence about this could be added to the text.]}
\label{fig:images}}
\end{figure*}

Figure \ref{fig:images} shows the $2''\times 2''$ target vicinity in the UVIS F475X and WFC F775W filters 
%\yus{The 
(the latter image is from the longer F775W observation of 2020 June~5).
%a combination of the images of 2020 June 5 and 2021 March 6
%two sets 
%where the second set image is aligned to the first one in accord to the pulsar p.m.
%\gp{\bf [was it indeed shifted by 22 mas (radio p.m.) or 48 mas (apparent optical shift)?]}
%which results in some blur of the combined image due to different orientations. May be it is reasonable to show here only the first set? Otherwise, potential referee can ask what is the reason of the blur as compared to the F475X image.}
In each of the images
we see only one 
%\yus{
firmly detected point source, %\yus{
presumably the pulsar counterpart. 
%{\bf [if we are sure that the smudge NNE of the pulsar in the F475X is not a real source, would it be possible/reasonable to remove it? otherwise we would have to describe it.]}
The nominal coordinates of this source in the pipeline-processed MAST 
images are RA = 09\h53\m09\fs3213 and Decl = $+$7\degs55\amin36\farcs673 in the F475X image, 
and RA = 09\h53\m09\fs3028 and Decl =  $+$7\degs55\amin36\farcs468 in the first F775W image; 
%\yus{It is natural to include here the 
 %centroid errors in brackets and note  that offsets are higher than these and pulsar coordinate uncertainties  ! }
they differ from the radio pulsar coordinates at these epochs %\yus{
(listed in Table~\ref{tab:psr-positions})  
by 
%265 mas and 283 mas,
0\farcs27 and 0\farcs28, respectively. These offsets are significantly larger than the 15 mas uncertainties in RA and Decl of the pulsar's radio position.
%the respective source centroid uncertainties of 5 and 8 mas
%\gp{[these are uncertainties of what? ``radial'' in the UVIS and WFC images? RA and Decl in one of the images?]}
%and the 15 mas uncertainty of the pulsar radio position.
%\gp{[in RA and Decl?]}
%The source centroiding uncertainties in the drizzled images are 5 mas in the F475X and 8 mas and 6 mas 
%in the first and second F775W images respectively. % in both coordinates.
Therefore, more accurate astrometric referencing of the HST images is required to firmly identify the optical source 
%with 
as the pulsar counterpart. 
%uncertainties of The source centroiding uncertainties in the drizzled images are 5 mas in the F475X and 8 mas and 6 mas 
%in the first and second F775W images respectively.% in both coordinates.  
%\gp{[Do we need this sentence, or at least the last words? I don't think that anybody would really doubt that this source is the psr counterpart.]}

%(see Table \ref{tab:psr-positions}).
%for F475X and F775W images.
%Such offets are roughly consistent with the error budget of the HST absolute astrometry of $\sim 0\farcs1$\footnote{   
%\gp{[Actually, they claim that the typical error in absolute astrometry is only about 0.1 arcsec nowadays, after they tied their GSC to Gaia (in 2017) -- 
%See \url{https://www.stsci.edu/files/live/sites/www/files/home/scientific-community/software/drizzlepac/_documents/drizzlepac-handbook.pdf} , Sec 4.5.}  
%Based on rom the comparison with the previously obtained images of the B0950 vicinity, it is clear
%To confirm 
%Therefore, we conclude
%that this source is the B0950 counterpart. 

%%%%%%%% Fig 1 %%%%%%%%%%%
%%%%%%%%%%%%%%%%%%%%%%%%
%%%%%%%% Table 3 %%%%%%%%
\begin{deluxetable*}{cccccccc}[t!]
\tablecaption{Optical and radio positions of PSR B0950$+$08 at different epochs
%and its p.m.\ components  
\label{tab:psr-positions}}
\tablecolumns{7}
\tablenum{2}
\tablewidth{0pt}
\tablehead{ 
\colhead{Visit } &
\colhead{Epoch } &\colhead{RA (J2000)} &\colhead{Decl (J2000)} & \colhead{RA (J2000)} & \colhead{Decl (J2000)} & \colhead{$\Delta$RA} & \colhead{$\Delta$Decl} \\ 
& \colhead{MJD} & \colhead{(optical)} & \colhead{(optical)}  & \colhead{(radio)} & \colhead{(radio)} & (mas) & (mas) }
\startdata
 & 51544 & \nodata & \nodata & 09\h53\m09\fs3071(10) & $+$7\degs55\amin36\farcs148(15) & \nodata & \nodata \\
1 &58984 & 09\h53\m09\fs3049(5) & $+$7\degs55\amin36\farcs761(7) & 09\h53\m09\fs3040(10) & $+$7\degs55\amin36\farcs748(15)& $13\pm 17$ & $13\pm 17$ \\
2 &59005 & 09\h53\m09\fs3051(8) & $+$7\degs55\amin36\farcs740(11) & 09\h53\m09\fs3039(10) & $+$7\degs55\amin36\farcs750(15)& $18\pm 19$ & $-10\pm 19$\\
3 & 59279 & 09\h53\m09\fs3043(5) & $+$7\degs55\amin36\farcs781(7) & 09\h53\m09\fs3040(10) & $+$7\degs55\amin36\farcs772(15)& $4\pm 17$ &  $9\pm 17$\\
\enddata
\tablecomments{
%\gp{
%Are these coordinates and the p.m. from astrometry of optical images? What do we want to say the reader by this table?}
%[Where the radio coordinates (and proper motion) are taken from? I see the POSEPOCH = MJD 46375 in the ATNF catalog.]} 
The reference radio coordinates of the pulsar (first line) are taken from \citet{Brisken2002} The radio coordinates at the epochs of the HST observations are calculated using the best-fit values of proper motion $\mu_\alpha = -2.09$ mas yr$^{-1}$,
%\pm 0.08$ mas yr$^{-1}$, 
$\mu_\delta = +29.46$
%\pm 0.07$ 
mas yr$^{-1}$, and parallax $\pi = 3.82$ mas \citep{Brisken2002}. 
%\gp{[What about parallax? It would decrease the predicted $\alpha$ by about 3 mas for May 15 and June 5 (about 1 mas in Mar 6).]}
Offsets of the optical positions with respect to the radio positions are given in two last columns.  The numbers in parentheses are the numerical values of the standard uncertainty
(at the 68\% level of confidence) referred to the corresponding last digits of the quoted result.
%\va{Centroiding uncertainties of the pulsar were taken into account in the uncertainties of optical positions.}
}
\end{deluxetable*}
%%%%%%%%%%%%%%%%%%%%%%%%%%

To precisely  measure the  International Celestial Reference System (ICRS)  coordinates of the point source,
%counterpart,
%at the epochs of our HST observations,  
%transformed the nominal UVIS and WFC coordinates to the most accurate 
we obtained astrometric solutions for our images using the Gaia EDR3 Catalog \citep{GaiaCollaboration2021} and the IRAF tasks {\tt imcentroid} and {\tt ccmap}.
%\gp{[In principle, we can omit the absolute astrometry part...]}
%to obtain astrometric solutions. 
%
In %each of 
the UVIS and the first WFC fields of view 
we found five unsaturated cataloged stars, two of them 
%within 
%in the UVIS field of view (FoV)
%of about $162\asec \times 162\asec$ 
%as well as within the WFC FoV.
%of about $202\asec \times 202\asec$. 
%Two of these stars 
are common 
%for 
to both fields, and only four catalogued stars in the second WFC field of view, three of them are common with the first WFC FOV. 
%The GAIA reference epoch is 2015.5, therefore we calculated 
To calculate the  coordinates of the reference stars at the epochs of the HST observations, 
%using 
we shifted their Gaia EDR3 positions, given  at the reference epoch 2016.0, 
%to the epochs of our HST observations 
using  the cataloged proper motion values.
%from the DR2 Catalog. 
%We used the {\sl Gaia} DR2 Catalog \citep{Lindegren2018} and the IRAF tasks {\tt imcentroid} and {\tt ccmap }
%to obtain astrometric solutions. 
%
%In each of the UVIS and WFC fields of view 
%we found five unsaturated catalogued stars; two of them 
%within 
%in the UVIS field of view (FoV)
%of about $162\asec \times 162\asec$ 
%as well as within the WFC FoV.
%of about $202\asec \times 202\asec$. 
%Two of these stars 
%are common 
%for 
%to both fields. 
%The GAIA reference epoch is 2015.5, therefore we calculated 
%To calculate the  coordinates of the reference stars, 
%at the epochs of the HST observations, 
%using 
%we shifted the Gaia DR2 positions, given  at the reference epoch 2015.5, to the epochs of our HST observations using  the catalogued proper motion values.
%from the DR2 Catalog. 
%\gp{Which tool was used for astrometry?} \va{ccmap}
%\yus{
Astrometric fits yielded   
%We got the 
the root-mean-square (rms) %astrometric fit 
residuals of 2.4 mas in the RA and 3.3 mas in the Decl 
%with 
for the F475X image, and
6.8 (4.0) mas and 4.4 (1.8) mas in the RA and Decl, respectively, 
for the first (second) F775W images. 
%\gp{Are these numbers for first F775W observation only?}
The rms centroiding uncertainties of the reference stars are $\lesssim$ 0.3 mas in the RA and Decl for all three images.
%for F475X, and 8.9 (10.2) mas in each coordinates for the first (second) F775W images. 
%\gp{[Do these centroiding uncertainties really matter once the rms residual for the frame transformation are so small?]
%}
The %\gp{average? typical?} 
rms uncertainties of the Gaia %coordinate errors 
positions of the reference objects %are
%\gp{[can we write `do not exceed'? (then `average' or `typical' are not needed.]} 
do not exceed 0.5 mas in both coordinates and within each field.
%\gp{[[Is it possible to write where this naumber was taken from?]}
The proper motion corrections have rms uncertainties of 4.9 mas in the RA and 4.3 mas in the Decl for the F475X image, and 5.3 (2.3) mas and 4.8 (1.7) mas in the RA and Decl, respectively, for the first (second) F775W observations. 
As a result, combining all the uncertainties in quadrature, we obtain 5 mas in both coordinates 
for astrometric referencing uncertainties of the F475X image. 
%Whereas for 
We got 9 (4) mas and 7 (3) mas in the RA and Decl, respectively for referencing of the first (second) F775W images. 
%RA and Decl.
%\gp{Again, are these uncertainties for the pulsar positions (as centroiding errors are different for different sources)?}

Using the Gaia-based astrometric solutions, we
measured the ICRS coordinates of the 
point source. They are listed in Table~\ref{tab:psr-positions}, where the coordinate errors account for the astrometric 
referencing uncertainties and  the source centroding uncertainties of 5 mas in the F475X image and 8 and 6 mas 
in the first and second F775W images, respectively.
%\gp{[Oleg's remark: What were the centroiding uncertainties of the pulsar in the HST images (it might be reasonable to put them in?).]}
%uncertainties of The source centroiding uncertainties in the drizzled images are 5 mas in the F475X and 8 mas and 6 mas 
%in the first and second F775W images respectively.% in both coordinates.   
%Were they taken into account in the 'optical' uncertainties in Table 3?]}
We compared them with predicted coordinates of the radio pulsar at the epochs of the HST observations,  
%point-like object which are given in 
%(Table~\ref{tab:psr-positions}), 
using the reference radio position and the proper motion and parallax provided by \citet{Brisken2002}.
%....  ... 3 positions and radio position...  ... maybe in a table?
%Thus, there remain no doubts that the point source is indeed the pulsar counterpart... 
%\gp{A couple of words that we were able to detect the pulsar shift during just one year...}
As seen from Table~\ref{tab:psr-positions},  %We see that 
the optical-radio offsets are in the range of 4--18 mas, within their statistical  %measurement %coordinate 
uncertainties.
%\yus{of the source and radio pulsar in the images. The offsets are about ten times smaller than the  pipeline 
%offset mentioned above. }
This demonstrates the high accuracy of the Gaia-based HST astrometry and firmly confirms that the detected point source is indeed the pulsar counterpart.

In the HST images we do not see any extended emission 
%associated with the pulsar. 
that could be interpreted as a pulsar wind nebula (PWN).
%Our high-resolution images do not confirm 
An extension of the pulsar image in the direction perpendicular 
to the pulsar's trajectory,
%was 
reported 
%by Zharikov et al (2002, 2004)
from the Subaru and VLT observations 
in the B band 
%\yus{ 
and suggested to be a PWN by \citep{Zharikov2002, Zharikov2004}, %Zharikov et al (2002, 2004),
%possible signature the pulsar wind nebula (PWN). From the HST data, we see that the apparent extension was likely 
might be due to 
%the presence 
%the contribution of 
%contamination from 
overlapping with a faint blue background source 
located 
%$\approx 
$0\farcs6$ south of the pulsar 
%and seen 
in the F475X image (i.e., 0\farcs2 west from the pulsar position in 2001, when the Subaru and VLT observations were taken, with seeing of 0\farcs7).
In the F775W image we see an extended object with a non-uniform brightness distribution, apparently a distant galaxy or a couple of galaxies, at about 0\farcs3--0\farcs5 north-northwest of the pulsar. Only a hint of it can be seen in the F475X image. This is the object O2 that \citet{Zharikov2004} detected in 2001 with the VLT in the I and R bands, when the pulsar was $0\farcs58$ farther south. Obviously, if the pulsar were observed now with a ground-based telescope without adaptive optics, it would not be be resolved from the extended source. 
%\yus{Its  closeness  to the pulsar was the main reason why we were not satisfied by the MAST pipeline products with a nominal 
%pixel size of 0\farcs05 and re-drizzled the images with the twice smaller scale increasing the spacial resolution. } 
%\gp{[not sure it is needed here. we shall see..]}

%\gp{[Maybe a couple of words on nonuniformity of the images??? --- I meant the apparent nonuniformities  
%the diagonal of the square of the F435X imagalonge, just to explain the choice of the bgd region and maybe to mention a possible tail. Perhaps not really needed here, could be mentioned below in the Photometry section.]
%}

\subsection{Photometry}
\label{phot}
%\gp{It would be good to explain here or above why we were not satisfied by the pipeline products, why we chose such a pixel size in the drizzled images.}
%The pixel scale in the ``drizzled'' images is 0\farcs025
%for both filters.
We measured the 
%\yus{pulsar counterpart} source
pulsar count rates 
%in the F475X and F775W filters, using 
in  apertures of 5.6 pix (0\farcs14) and 5 pix (0\farcs125) radii in F475X and F775W filters, respectively. 
To determine the encircled energy fraction $\phi_E$ in these apertures, we measured $\phi_E$ as a function of aperture radius for 2 relatively bright, unsaturated stars in the F475X image and 5
%For both filters we found encircled energy fraction as a function of aperture using 
%5 relatively bright unsaturated isolated 
stars in the F775W image.
%and 2 stars in the F475X image. 
The  obtained $\phi_E = 0.80$  and 0.73, respectively, are close to their nominal values for these detectors\footnote{
%quantities comparable with the modelled %ones for F775W filter with ACS/WFC\footnote{
%$\phi_E=0.XX$ and 0.YY, respectively;
See \cite{Deustua2017} and \url{ https://www.stsci.edu/hst/instrumentation/acs/data-analysis/aperture-corrections}.} 
%and F475X filter with WFC3/UVIS \citep{Deustua2017}
for the chosen aperture sizes.

In both the F475X and F775W fields we measured the background in parts of annuli 
%two $90^\circ$ segments of annulus 
with 10 pix (0\farcs25) and 30 pix (0\farcs75) inner and outer radii (see Figure \ref{fig:images}).
In the F475X image we extracted background counts from two $90^\circ$ segments east and west of the pulsar to avoid an apparent background enhancements
(some of them may be very faint sources)
%along the pulsar path \yus{(
along the vertical  diagonal of the square of the F475X image.
In the F775W images, the background was extracted  from the half-annulus 
%with the same inner and outer radii
(the northern half of the annulus was excluded because it contains the extended source).
%of 10 pix 
%and 30 pix inner and outer radii. respectively. 

Since the numbers of counts in the small 
pixels of the drizzled images are not statistically independent, the usual approach to noise evaluation, based on mean and standard 
deviation of counts (or count rates) per pixel\footnote{See, e.g., Equation (25) in \citet{MerlineHowell1995}}, 
%Merline \& Howell (1995). \gp{[Add to the bibliography]}
yields underestimated uncertainties of source count rate\footnote{See, e.g., Appendix in \citet{Casertano2000} 
%Casertano et al.\ (2000) \gp{[add to the bibliography]} 
and Section 3.3 in the DrizzlePac Handbook \url{https://www.stsci.edu/files/live/sites/www/files/home/scientific-community/software/drizzlepac/_documents/drizzlepac-handbook.pdf}.}. Therefore, we used the ``empty aperture'' approach (see Section 3.4 in \citealt{Skelton2014}). In this approach background 
%counts 
count rates are measured in multiple apertures 
of the same size as the source aperture, randomly placed 
within the background region. The mean
$\overline{C}_{\rm bgd}$ and the %standard deviation 
variance $\sigma_{C_{\rm bgd}}^2$ of 
%the number of counts 
count rates per aperture yield the %direct (and 
most reliable estimates for the net 
%number 
%of source counts $N_s$ 
source count rate and its uncertainty  $\sigma_s$:
\begin{equation}
    C_s = C_{\rm tot} -\overline{C}_{\rm bgd},\quad\quad \sigma_s = \left(\sigma_{C_{\rm bgd}}^2 + C_st_{\rm exp}^{-1}  %\sigma_{C_{\rm bgd}}^2
    \right)^{1/2},
\end{equation}
where $C_{\rm tot}$ is the total %number of counts 
count rate in the source aperture, and $t_{\rm exp}$ is the exposure time. Notice that the directly measured quantity $\sigma_{C_{\rm bgd}}^2$ includes all possible sources of background noise, i.e., the sky contribution as well as various instrumental and processing contributions. An advantage of this empirical approach is its simplicity and minimum of assumptions involved.

In our case, for each of the images we used a set of 1000 circular 
%regions of size 
apertures with the same radii as the source apertures, $r_{\rm extr}=0\farcs14$ and 0\farcs125 for the F475X and F775W images, respectively.
%\sout{filter and $r_{\rm extr}$=0\farcs125 for F775W filter, randomly selected in the background region to estimate the background standard deviation $\sigma_{bgd}$.}
The results 
%(in terms of count rates)
for each of the 3 visits 
%  and for the sum of the images obtained in 2 visits in which the F775W filter was used 
are presented in
Table~\ref{tab:photometry}.

%\gp{I am trying to reproduce 6th line of Table 2:\\
%Source aperture area $A_s = 78.54$ pix = 0.0491 arcsec$^2$.\\
%Background aperture area $A_b=1256.5$ pix = 0.785 arcsec$^2$.\\
%$A_s/A_b=0.0625$.\\
%Background count rate in the source aperture $C_b = 0.34\times 78.5
%=26.7$ cnt/ks.\\
%Source count rate in the source aperture $C_s = C_t-C_b = 196.3$.\\
%The variance of the source count rate in the ``standard approach'' is 
%\begin{equation}
%    \sigma_{C_s}^2 = C_s t_{\rm exp}^{-1} + \sigma_{b,pix}^2 A_s  (1+A_s/A_b),
%\end{equation}
%where $\sigma_{b,pix}^2$ is the %standard deviation 
%variance of the bgd count rate per pixel ($\sigma_{b,pix}^2 = 1.1^2$ cnt$^2$/ks$^2$/pix in Table 2), and $t_{\rm exp} = 6.16$ %ks is the exposure time.  According to this formula, $\sigma_{C_s}^2 = 196.3/6.16 + 1.21\times 78.5\times 1.0625 = 132.8$ %(cnt/ks)$^2$ = $(11.5\,{\rm cnt/ks})^2$, whereas $C_s = 196\pm 17$ cnt/ks in Table 2. Why such a difference?
%}

To convert the source count rate, corrected for the finite aperture size, into  mean flux density, we used the inverse sensitivity header keyword 
%PHOTFNU 
{\tt photfnu} (${\cal P}_\nu$ in Table \ref{tab:photometry}) for the F475X image. 
%At the same time, for 
For the F775W images we calculated %the count-rate-to-flux conversion factor 
${\cal P}_\nu$, using the header 
keywords %PHOTFLAM and PHOTPLAM. 
{\tt photflam} and {\tt photplam}. The resulting source fluxes are presented in first 3 lines of Table \ref{tab:photometry} for each observation.

%\yus{To increase the signal to noise ratio in the F775W filter, we combined the data from the two visits. We found %that the intensity peak of the counterpart in the 2021 image is shifted  by about one pixel toward 
%north in respect to that in the 2020 image. This is roughly compatible with the expected pulsar 
%proper motion shift of 20 mas between the two visits. We thus used the brightest pixels of the counterpart in both images as reference points to align the second image to the first one before combining them.}
%\gp{[I don't think we need these details unless we want to discuss the proper motion in 9 month (which is not useful)].}
In fourth line of Table \ref{tab:photometry} we provide photometry results for the sum of the two F775W images, %with 
%using the brightest pulsar pixels for alignment.
aligned on the pulsar position. 
%The counterpart 
The pulsar's mean flux density,
$\overline{f}_\nu = 52.9\pm 6.9$ nJy, estimated 
from the combined  image 
%is  presented in  Table~\ref{tab:photometry}. It
is in excellent agreement with     the weighted mean flux density, $\overline{f}_\nu = 52.2\pm 6.8$ nJy, of the flux densities measured in the separate visits.
%\gp{[Frankly, there were no need to combine the images for the sake of photometry -- combining just can add an additional uncertainty]}
%with this filter. % is in good agreement with the flux density estimated in the summed image also  presented in  Table~\ref{tab:photometry}. 
%[[{\bf do we really need the flux from the summed image? 
%Did I retain right lines in Table 2?}} ]] 
%\va{I used the flux from the summed image in the spectral fit. }
%Images from the separate visits were combined with accounting for the pulsar proper motion. We corrected WCS of the second image to align pulsar position to the first image. 
%\gp{Then we have to describe how that image was produced, with account for the fact the the psr has moved by 20 mas, about one pixel.}

We see that while the F475X flux density is in good agreement with the Subaru and VLT 
measurements in the B band, the F775W flux is considerably lower than the I and R fluxes measured with the VLT (see Section~\ref{introduction} and \citealt{Zharikov2004}). The reason for this discrepancy is the relatively poor angular resolution of the VLT observations, which 
%prevented an accurate
hampered correction for contamination from the nearby extended object north of the pulsar. 

%\yus{
In  Table~\ref{tab:photometry},  we also include 
%also  
%the 
%corrected 
FUV photometry results 
%of the pulsar counterpart
%obtained by 
\citep{Pavlov2017} 
%with the ACS/SBC detector  in  the F140LP and F125LP filters. They are corrected 
corrected with account for the new ACS/SBC calibration 
%results 
presented by 
\citet{Avila2019}.
The lower values of the corrected inverse sensitivity
%which are accounted for in the count-rate-flux-density   
%conversion factors  
${\cal P}_\nu$
%.  The resulting flux densities are about 20-30 percent 
resulted in lower FUV flux densities than those published by \citet{Pavlov2017}, by 29\% and 24\% for the F125LP and F140LP filters, respectively.  
%}

%\gp{Shouldn't we add a line in Table 2 and a paragraph in the text about the FUV exposures? I think we should.}
%\gp{Do I understand correctly that you did not re-analyze the FUV data but just used lower ${\cal P}_\nu$, taken from the new file headers?} \va{Yes}

%%%%%%%%%%%%%%%%%%%%%%%%%%%%%%%%%%%%%%%%%%%%%%%%%%%%%%%%%%%%%%%%%%%%%%%%%%%%%%%%%%%
\begin{deluxetable*}{cccccccccccc}[ht]
\tablecaption{Photometry of the pulsar counterpart \label{tab:photometry}}
%\tablecolumns{8}
\tablenum{3}
%\tablewidth{Opt}
\tablehead{
%\colhead{Image}
%\colhead{Filter} &
%\colhead{$t_{\rm exp}$} &
\colhead{Filter} & \colhead{Visit} & 
\colhead{$t_{\rm exp}$} & \colhead{$r_{\rm extr}$} & \colhead{$\phi_{E}$} & \colhead{$C_{\rm tot}$} & \colhead{$C_{\rm bgd}$}
%$\overline{C}_{\rm bgd} \pm \sigma_{C_{\rm bgd}}$} 
& $N_{b}$ & \colhead{$C_{s}$} & \colhead{$C_{\rm corr}$} & \colhead{${\cal P}_{\nu}$} &
%\colhead{${\cal P}_{\nu}$} & 
\colhead{$\overline{f}_{\nu}$}\\
%\colhead{ } & \colhead{(ks)} & 
& & \colhead{(s)} & \colhead{(arcsec)} & \colhead{(\%)} 
& \colhead{(cnts\,ks$^{-1}$)} & \colhead{(cnts\,ks$^{-1}$)} & \colhead{(cnts)} & \colhead{(cnts\,ks$^{-1}$)} & \colhead{(cnts\,ks$^{-1}$)} & \colhead{(nJy\,ks\,cnts$^{-1}$)} & \colhead{(nJy)}
}
\startdata
%F475X & 4936 & 0.14 & 80 & 417 & 1.2 & 3.9 & $371 \pm 27$ & $464 \pm 34$ & 0.123 & $57.0 \pm 4.2$\\
%F475X & 4936 & 0.14 & 80 & 426 & 81 & 44 & $345 \pm 45.6$ & $431 \pm 57$ & 0.123 & $53.0 \pm 7.0$\\
%F475X & 4936 & 0.14 & 80 & 426 & 41 & 43 & $385 \pm 45$ & $481 \pm 56$ & 0.123 & $59.0 \pm 6.9$\\
F475X & 1 & 2420 & 0.14 & 80 & 426 & $38 \pm 36$ & ... & $388 \pm 38$ & $485 \pm 48$ & 0.123 & $59.6 \pm 5.9$\\
%F475X-10rnd & 4936 & 0.14 & 80 & 426 & 59 & 49 & $367 \pm 51$ & $459 \pm 64$ & 0.123 & $56.4 \pm 7.8$\\
%F475X-10 & 4936 & 0.14 & 80 & 426 & 50 & 31 & $376 \pm 39$ & $470 \pm 41$ & 0.123 & $57.8 \pm 5.1$\\
%F775W & 7693 & 0.125 & 73 & 272 & 0.85 & 4.9 & $255\pm 30$ & $349 \pm 41$ & 0.197 & $68.8\pm 8.1$\\
F775W & 2 &  3726 & 0.125 & 73 & 224 & $32 \pm 33$ & ... & $192\pm 34$ & $263 \pm 47$ & 0.197 & $51.8\pm 9.2$\\
F775W & 3 &  2434 & 0.125 & 73 & 222 & $22 \pm 36$ & ... & $200\pm 38$ & $274 \pm 51$ & 0.197 & $54.0\pm 10.1$\\
 F775W & 2+3 & 6160 & 0.125 & 73 & 223 & $27 \pm 25$ & ... & $196\pm 25$ & $268 \pm 35$ & 0.197 & $52.9\pm 6.9$\\
%{\bf F775W-5} & 7693 & 0.125 & 73 & 223 & 0.34 & 1.1 & $196\pm 11$ & $268 \pm 16$ & 0.197 & $52.9\pm 3.1$
\hline 
F125LP & ... & 5528 & 0.25 & 66  & 69 & ... & 1859 & $60.8 \pm 3.6$ & $92.0 \pm 5.4$ & 0.847 & $77.9 \pm 4.6$ \\
F140LP &... & 1774 & 0.25 & 67 & 35 & ... & 642 & $25.9 \pm 4.5$ & $39.0 \pm 6.7$ & 1.608 & $62.7 \pm 10.8$\\
\enddata
%\tablenotetext{a}{At exposure start.}
\tablecomments{%First column provides the filter name and the number of dither point used. 
$\phi_E$ is the fraction of source counts in the aperture with radius $r_{\rm extr}$,
$C_{\rm tot}$ is the total count rate in the source aperture,
$C_{\rm bgd}=\overline{C}_{\rm bgd}\pm\sigma_{C_{\rm bgd}}$, 
%where 
$\overline{C}_{\rm bgd}$ and $\sigma_{C_{\rm bgd}}$ are the mean and standard deviation of background measurements,
$N_{b}$ is the number of %background 
counts in the FUV background apertures,
$C_{s}$ is the net source count rate, its error is estimated as 
$[(\sigma_{C{\rm bgd}}^2 
%t_{\rm exp})^2 
+ C_{s} t_{\rm exp}^{-1}]^{1/2}$
%/t_{\rm exp}$ 
for optical images and 
$[C_{\rm tot} t_{\rm exp} + (A_{s} / A_{b})^2 N_{b}]^{1/2}t_{\rm exp}^{-1}$ for FUV images, where $A_{s}$ = 0.196 arcsec$^{2}$ 
%source aperture 
and 
$A_{b}$ = 7.854 arcsec$^{2}$ are the areas of the source and background apertures
%aperture background annulus
\citep{Pavlov2017}, 
$C_{\rm corr}$ is the aperture-corrected source count rate, $\overline{f}_\nu = C_{\rm corr} {\cal P}_\nu$ is the mean flux density, 
%at the pivot wavelength $\lambda_{\rm piv}$ 
%\gp{[not true, it is just a `mean flux density'?]},
and ${\cal P}_\nu$ is the count rate-to-flux conversion factor.
%\gp{
%We could combine 5th and 6th columns in one.
%After inclusion of the numbers of counts for the FUV bands, the table became ugly. Shouldn't we transform them to count rates in a clever way? Or %maybe transform count rates to numbers of counts for the optical bands?}
}
\end{deluxetable*}
%%%%%%%%%%%%%%%%%%%%%%%%%%%%%%%%%%%%%%%%%%%%%%%%%

\subsection{Spectral fits}
\label{sec:spectralfits}

%We used the results of our photometry to fit the broad band spectrum of B0950.

Following 
%Similar to 
\citet{Pavlov2017}, we fit the optical-UV spectrum with the PL+BB model
\begin{equation}
    f_\nu^{\rm mod} =\left[f_{\nu_0} \left(\frac{\nu}{\nu_0}\right)^\alpha + \frac{R_\infty^2}{d^2} \pi B_\nu(T_\infty)\right]\times 10^{-0.4 A_\nu},
\label{eq:PL+BB}
\end{equation}
where $\nu_0$ is the reference frequency for the PL spectrum (we chose $\nu_0 = 1\times 10^{15}$ Hz, which corresponds to $\lambda_0 = 3000$ \AA),
$d=262 d_{262}$ pc is the distance, 
%\yus{
$T_\infty=T (1+z)^{-1}$
%~K  
and  $R_\infty=R(1+z)$
%~km  
are the NS 
temperature and radius
%, respectively, 
as seen by a distant observer, $z=[1-2.953 %M_{NS}
(M/M_\odot)(1\,{\rm km}/R)]^{-1/2} - 1$
%R_{NS}(km)]^{-1/2}$ 
is the gravitational redshift, 
%%
%$R_\infty = 15 R_{15} $ km and $T_\infty=10^4 T_4$ K are the NS radius and brightness temperature as seen by a distant observer, 
%
%$d=262 d_{262}$ pc is the distance, $T\infty=10^5 T_5$ K is the 
%
and $B_\nu(T_\infty)$ is the Planck function. 
Below we use the notations $R_\infty = 15 R_{15} $ km and $T_\infty=10^4 T_4$ K; at a 
%canonical 
NS mass 
$M=1.4M_{\odot}$, the apparent radius  
$R_{\infty}=15$~km corresponds to an intrinsic (circumferential) NS radius $R=12.2$~km.
%} 
The extinction coefficient $A_\nu$ is proportional to color excess $E(B-V)$. According to \citet{Pavlov2017}, the most plausible range of the color excess for B0950 is $E(B-V) = 0.01$--0.03, but higher values, perhaps up to 0.06, are not excluded. We will consider the range $E(B-V)=0.01$--0.06 and use \citet{Cardelli1989} to connect $A_\nu$ with $E(B-V)$.

To fit the model given by Equation (\ref{eq:PL+BB}) to the measured broad-band spectrum, we vary the model parameters to minimize the following $\chi^2$ statistic
\begin{equation}
    \chi^2 = \sum_i \frac{\left(\overline{f}_{\nu,i} - \overline{f}_{\nu,i}^{\rm mod}\right)^2}{\sigma_{f_{\nu,i}}^2}\,,
\end{equation}
where $\overline{f}_{\nu,i}$ and $\sigma_{f_{\nu,i}}$ are the 
%count rate 
mean flux density and its error in the $i$-th filter 
%(denoted as $C_{\rm corr}$ and $\sigma_{C_{\rm corr}}$
(given in the last column of Table \ref{tab:photometry}), 
\begin{equation}
    \overline{f}_{\nu,i}^{\rm mod}= \left[\int f_\nu^{\rm mod} \nu^{-1} T_i(\nu)\,d\nu\right]
    \left[\int T_i(\nu) \nu^{-1}\,d\nu\right]^{-1} ,
\end{equation}    
and $T_i(\nu)$ is the passband throughput for the $i$-th filter as a function of frequency. 
%{\bf [CORRECT?]} \va{Yes}
In our fits we included all the HST observations available (5 filters) and the Subaru observation in the B filter as it was not contaminated by the red extended object north of the pulsar. The throughputs for the HST and Subaru B filters 
%\gp{[what about the B filter throughput?]}
were taken from the SVO filter profile service\footnote{\url{http://svo2.cab.inta-csic.es/svo/theory/fps3/index.php?mode=browse&gname=HST&gname2=ACS_WFC&asttype=}}.

\begin{figure*}[t]
\begin{minipage}[h]{0.5\linewidth}
\center{\includegraphics[width=1\linewidth]{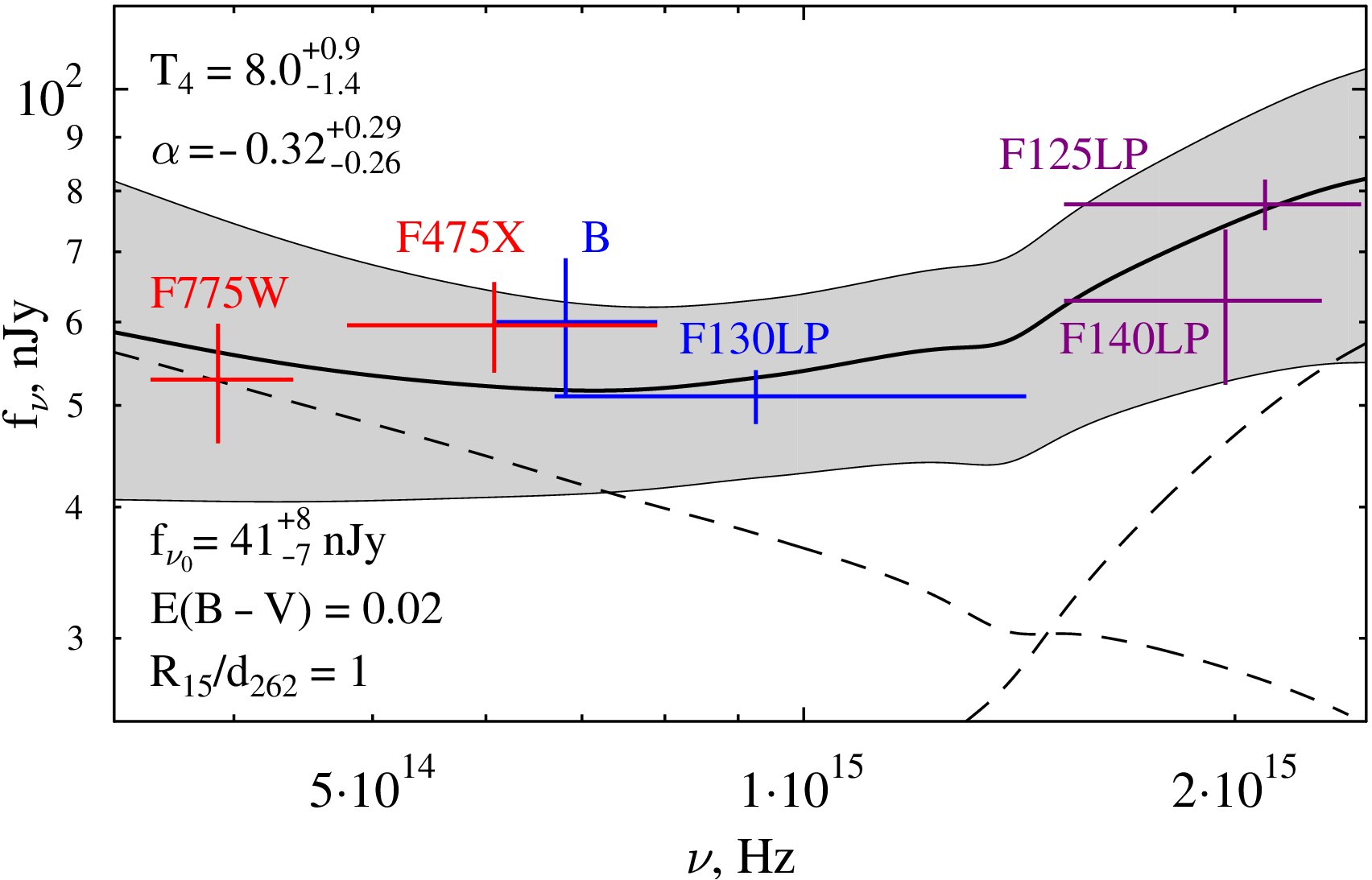}}
\end{minipage}
\hfill
\begin{minipage}[h]{0.5\linewidth}
\center{\includegraphics[width=1\linewidth]{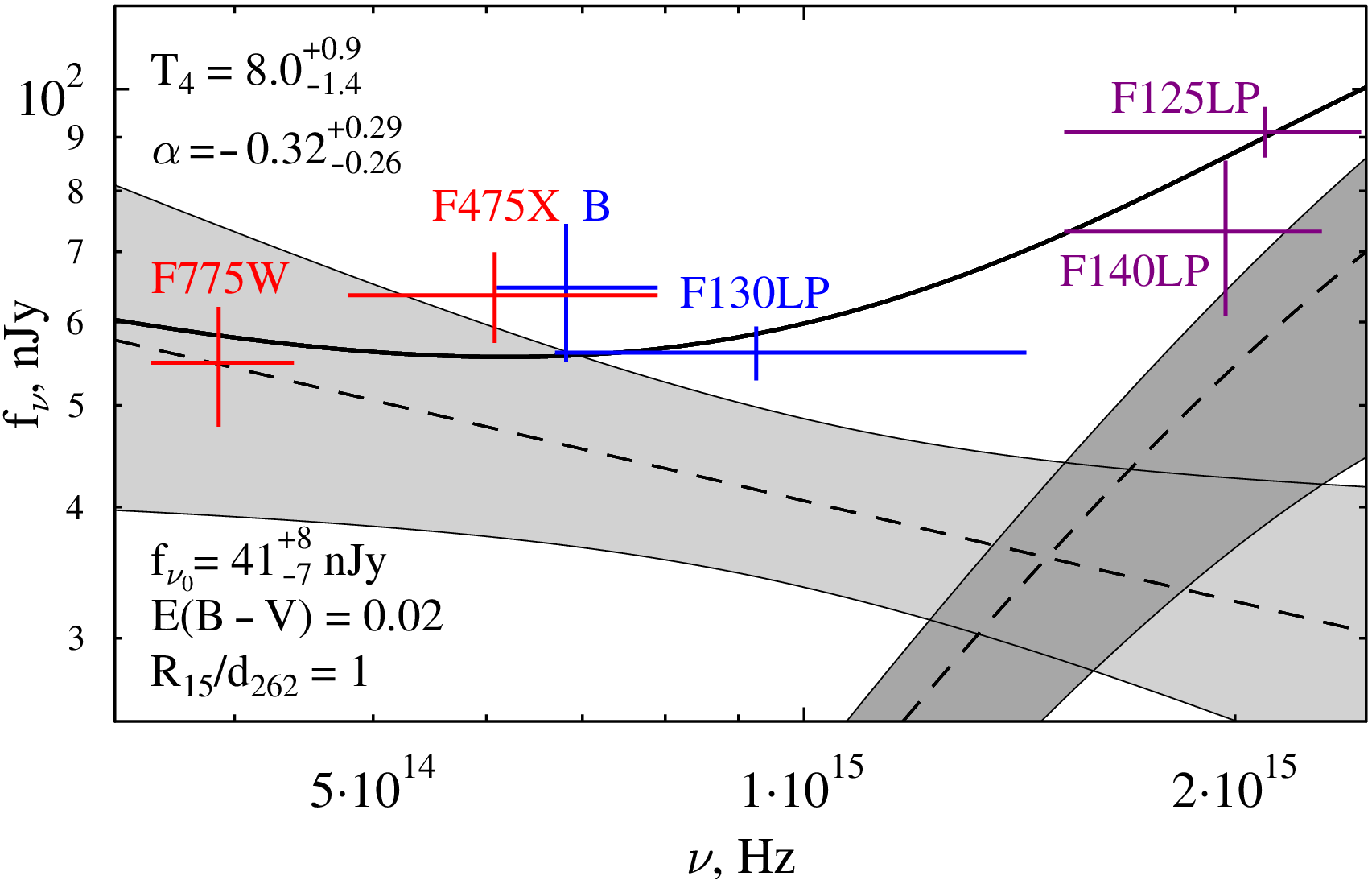}}
\end{minipage}
%\begin{center}
%\includegraphics[scale=0.19,angle=0]
%{0950-Best002-15-13.PNG} 
\caption{
%\yus{
%Unabsorbed ?}\va{absorbed}
%Broadband 
Observed (left) and dereddenned (right) optical-UV spectra  of the pulsar counterpart.
%for $E(B-V)=0.02$.
The thick solid lines show the best fit for the HST and Subaru (B) 
data with the 
%absorbed 
PL+BB model at fixed $E(B-V)=0.02$ and $R_{15}/d_{262} = 1$. 
%\yus{D
The dashed lines show the best-fit spectra of the PL and BB components. 
The shaded areas demonstrate fit uncertainties, at the 68\% confidence level, for the components (right panel) and 
their sum (left panel).
%\gp{[would it be possible to make the best-fit solid lines thicker, or the boundaries of the shaded areas thinner?]}
%\gp{[would it be reasonable to show fit uncertainties, maybe for the components? maybe in an additional panel with the unabsorbed spectra?]}
%\gp{
%$R_\infty = 15$ km seems more realistic nowadays (but we still can use 12 km here for comparison?)?
%Is it possible to put best-fit $\alpha$, $T_\infty$ and $f_{\nu,0}$ in the figure frame, maybe with errors? 
%and indicate 
%Is it possible to increase the y/x aspect ratio (stretch the frame along the y axis)? 
%}
\label{fig:fit}}
%\end{center}
\end{figure*}

The model given by Equation (\ref{eq:PL+BB}) has 5 parameters. Because of the few data points and a correlation between the parameters, we chose to fit 3 of them: $T_\infty$, $\alpha$ and $f_{\nu_0}$ at fixed values of $E(B-V)$ and $R_\infty/d = 
%15 km/262 pc 
1.86\times 10^{-15} R_{15}/d_{262}$ (cf.\ \citealt{Pavlov2017}). An example of such a fit at plausible values $E(B-V)=0.02$ and $R_{15}/d_{262}=1$ is shown in Figure \ref{fig:fit}.
The best-fit parameters are\footnote{Here and below the uncertainties of best-fit parameters are given at the 68\% confidence level for one parameter of interest unless noted otherwise.} $T_4 = 8.0^{+0.9}_{-1.4}$, 
%[($T_4 = 
%($8.0^{+1.4}_{-2.8}$ 
%and $8.0^{+2.0}_{-8.0}$ 
%for the 90\% 
%and 99\% 
%confidence level), 
%respectively) -- move below?]
%and 
$\alpha=-0.32^{+0.29}_{-0.26}$, and $f_{\nu_0}= 41^{+8}_{-7}$ nJy.
With the reduced  
$\chi_{\rm min}^2=1.46$ for 3 degrees of freedom, the fit is not perfect 
%-- the %minimum 
%reduced  
%$\chi_{\rm min}^2=1.46$ for 3 degrees of freedom 
(perhaps because there are systematic errors unaccounted for), but it is substantially better than the fit obtained by \citet{Pavlov2017}, with the reduced
%$\chi_{\rm min}^2 = 10.6$ for 5 degrees of freedom 
$\chi_{\rm min}^2 = 3.8$ for 4 degrees of freedom (for $E(B-V)=0.02$, $R_{15}/d_{262}= 0.8$).

%%%%%%%%%%%%%%%%%%%%%%% Figure 2 %%%%%%%%%%%%%%%%%%%%%%%%%%%%%%%%
%%%%%%%%%%%%%%%%%%%%%%%%%%%%%%%%%%%%%%%%%%%%%%%%
%With the reduced  
%$\chi_{\rm min}^2=1.46$ for 3 degrees of freedom, the best fit is not perfect 
%-- the %minimum 
%reduced  
%$\chi_{\rm min}^2=1.46$ for 3 degrees of freedom 
%(perhaps because there are systematic errors unaccounted for), but it is substantially better than the fit obtained by \citet{Pavlov2017}, with the reduced
%$\chi_{\rm min}^2 = 10.6$ for 5 degrees of freedom 
%$\chi_{\rm min}^2 = 3.8$ for 4 degrees of freedom (for $E(B-V)=0.02$, $R_{15}/d_{262}= 0.8$).
%in \citet{Pavlov2017}
%$\chi_{\rm min}^2 = 10.6$ was for PL model in P+17 
%\gp{[I don't understand now where this 10.6 came from. The non-reduced chi-square was 15.2, wasn't it?]}. 
%The best-fit parameters are\footnote{Here and below the uncertainties of best-fit parameters are given at the 68\% confidence level for one parameter of interest unless noted otherwise.} $T_4 = 8.0^{+0.9}_{-1.4}$, 
%[($T_4 = 
%($8.0^{+1.4}_{-2.8}$ 
%and $8.0^{+2.0}_{-8.0}$ 
%for the 90\% 
%and 99\% 
%confidence level), 
%respectively) -- move below?]
%and 
%$\alpha=-0.32^{+0.29}_{-0.26}$, and $f_{\nu_0}= 41^{+8}_{-7}$ nJy.
%\gp{[One could ask why the 90\% errors are given for $T_4$ only.]}
%Note that the lower bound on temperature becomes very low at high confidence level values; for instance uncertainty At a higher confidence level the 

For the direct comparison with \citet{Pavlov2017}, we also fit the model with $R_{15}/d_{262} = 0.8$. 
%\sout{\yus{($R=8.7$~km at $d_{262}=1$ and $M=1.4M_{\odot}$)}}.
%\gp{[not useful info.]}
%for comparison with \citet{Pavlov2017}.
Although the best-fit optical-FUV spectrum (reduced $\chi^2_{\rm min}=1.39$) is qualitatively similar to that shown in Figure 3 of 
Pavlov et al.\ (2017), 
%P+17, 
the fitting parameters are quite different: $T_4 = 10.1^{+1.5}_{-2.0}$, 
%[($T_4 = 
%($10.1^{+2.2}_{-4.2}$ 
%and $10.1^{+3.1}_{-10.1}$ 
%for the 90\% 
%and 99\% 
%confidence level)
%respectively) - can be omitted?]
$\alpha=-0.28^{+0.28}_{-0.26}$
versus $T_4=16.8\pm 2.6$, 
%\gp{was the new fit obtained for the same $R_\infty/d$? Apparently not. For comparison it would be perhaps more reasonable to use the same ratio as P+2017}
%and $\alpha=-0.28^{+0.28}_{-0.26}$ versus 
$\alpha = -1.17\pm 0.43$. 
The lower temperature and more gradual slope of
%\gp{ [less negative $\alpha$ is not a steepening of the slope, it is a more gradual slope]} 
of the  PL component 
%\yus{with the frequency} 
%\old{spectrum} 
are caused  
by lower values for the FUV flux due to the corrected calibration of the ACS/SBC sensitivity, and by
the lower optical flux, which is not distorted by
%contaminated by
%values in due to more accurate subtraction of 
a contribution from the nearby extended source, unlike the previous ground-based observations in red filters.
%, as well as by lower values for the FUV flux due to the better calibration of the ACS/SBC sensitivity.

%%%%%%%%%%%%%%%%%%%%%%% Figure 3 %%%%%%%%%%%%%%%%%%%%%%%%%%%%%%%%
\begin{figure}[ht]
\begin{center}
\includegraphics[scale=0.198,angle=0]
{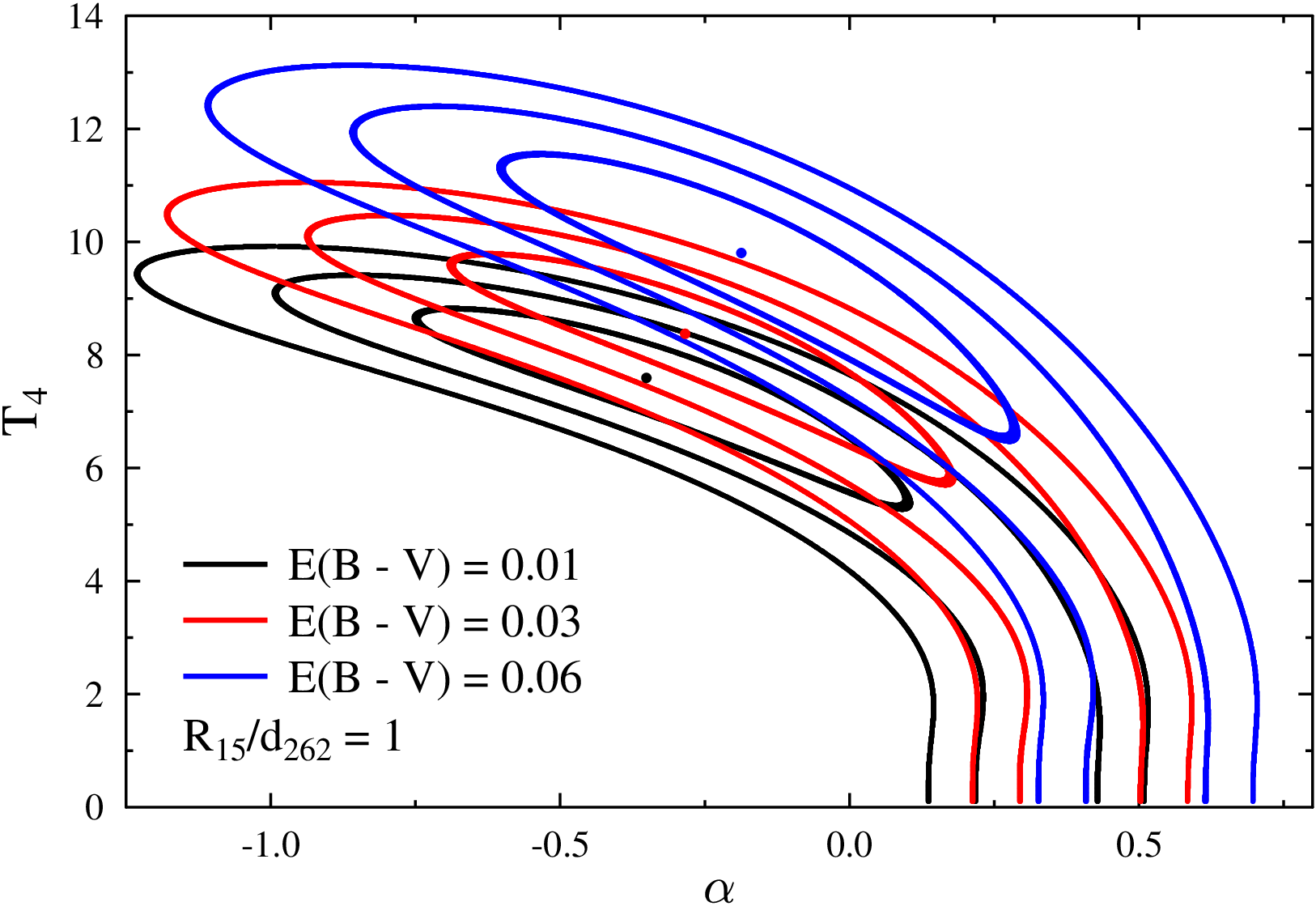} 
\caption{Confidence contours in the $\alpha$--$T_\infty$  plane
%gp{we should either use $T_\infty$ or $T_5$ (or $T_4$ defined wrt $T_\infty$ in the text.}
for $R_{15}/d_{262} = 1$ and $E(B-V)=0.01$, 0.03 and 0.06 (black, red and blue lines, respectively). %$E(B-V)=0.03$ 
%0.03 (red lines) 
%and 
%$E(B-V)=0.06$ 
%0.06 (blue lines). 
The contours 
%are lines of constant values of $\chi^{2}_{\rm min} + \Delta \chi^{2}$, where 
%$\chi^{2}_{\rm min} = 4.40$ 
%(non-reduced) for E(B - V) = 0.01 
%\gp{$ = 4.38$ (non-reduced!)} \va{4.38 for E(B - V) = 0.02} 
%$\chi^2_{\rm min}$ is the minimum value of non-reduced $\chi^{2}$ at given $E(B-V)$, and $\Delta \chi^{2}$ = 2.30, 6.17 and 11.8 for 
are plotted for the 68.3 \%, 95.5 \% and 99.7 \% confidence levels %respectively 
(for two parameters of interest).
%\gp{[shouldn't $\chi^2_{\rm min}$ be different for different $E(B-V)$?]} \va{$\chi^{2}_{\rm min} = 4.40$ for E(B - V) = 0.01; $\chi^{2}_{\rm min} = 4.38$ for E(B - V) = 0.02 ...}
The third parameter $f_{\rm \nu_0}$ was varied to minimize $\chi^{2}$ at each point of the $\alpha$--$T_\infty$ plane. 
%\gp{[Is it possible to replace the colored squares by color lines?]} \va{I will try}\gp{[and put $R_{15}/d_{262=1}$ in the figure panel?]}
\label{fig:T-alpha-contours}}
\end{center}
\end{figure}
%%%%%%%%%%%%%%%%%%%%%%%%%%%%%%%%%%%%%%%%%%%%%%%%

%%%%%%%%%%%%%%%%%%%%%%% Figure 4 %%%%%%%%%%%%%%%%%%%%%%%%%%%%%%%%
\begin{figure}[ht]
\begin{center}
\includegraphics[scale=0.195,angle=0]
{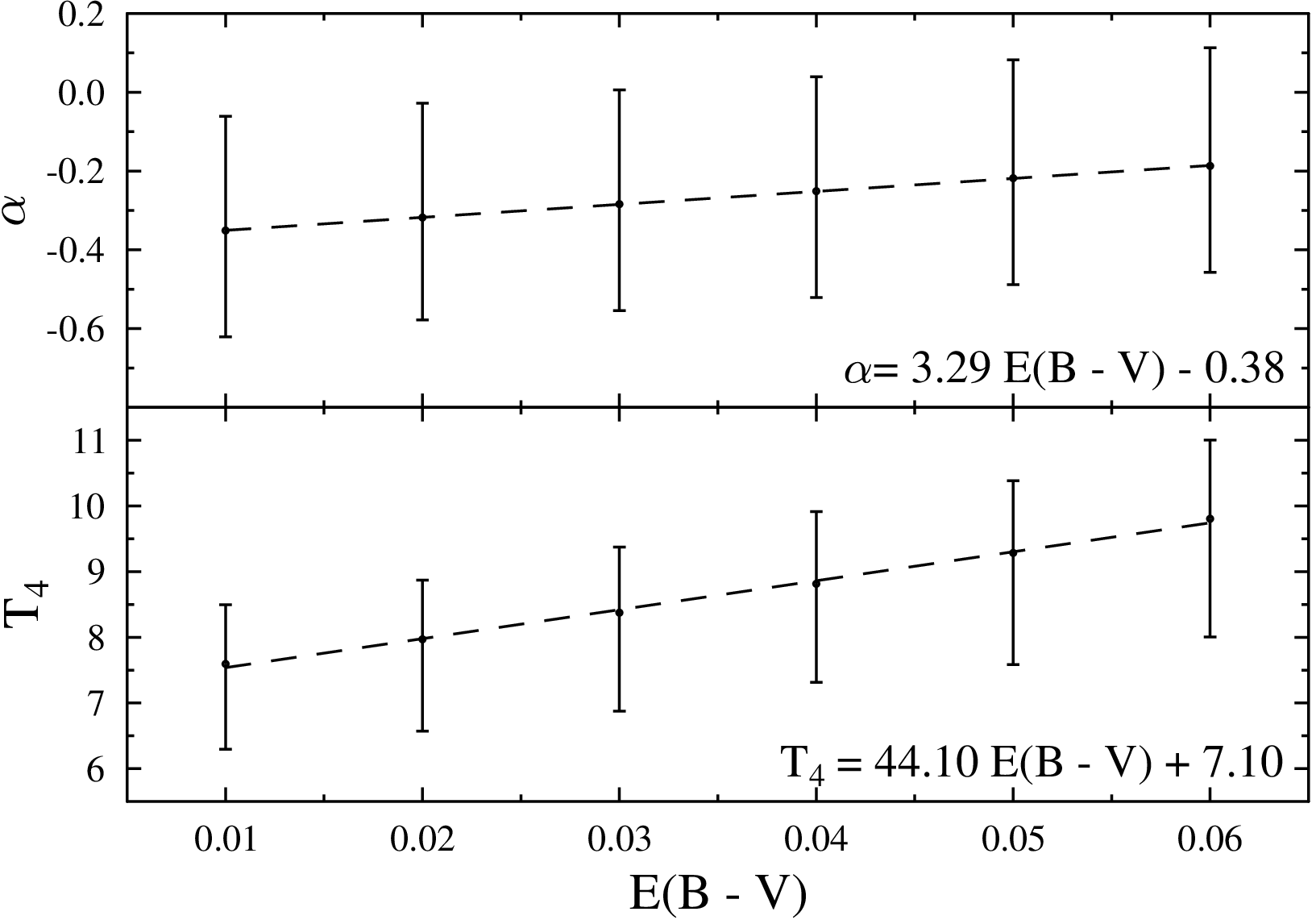} 
\caption{Dependencies of the main fitting parameters on color excess for $R_{15}/d_{262} = 1$. The uncertainties are at the 68\% confidence level.
%\gp{[Are the uncertainties at the 68\% level?]}
The equations in the panels are linear approximations of the dependencies. 
\label{fig:T-alpha-dependencies}}
\end{center}
\end{figure}
%%%%%%%%%%%%%%%%%%%%%%%%%%%%%%%%%%%%%%%%%%%%%%%%

The uncertainties of the fitting parameters grow with 
%increasing 
raising the confidence level. 
%In particular, 
Moreover, the lower bound on temperature tends to zero at a sufficiently high confidence level. For instance, 
in the above-considered example of $E(B-V)=0.02$, $R_{15}/d_{262}=1$, we obtain $T_4 =8.0^{+1.4}_{-2.8}$ at the 90\% confidence, but $T_4=8.0^{+2.0}_{-8.0}$ at the 99\% confidence.
%For $R_{15}/d_{262}=0.8$ the best-fit values and upper bounds are slightly higher, but we see the same trend:
%$T_4 = 
%10.1^{+2.2}_{-4.2}$ 
%and $10.1^{+3.1}_{-10.1}$ 
%and $10^{+3}_{-10}$
%at the 90\% 
%and 99\% 
%confidence levels.
%\gp{[I would remove this sentence -- it does not add anything to the qualitative picture.]}
%, for $R_{15}/d_{262}=0.8$).
%respectively.
%higher than \gp{???\%}
%The uncertainties of the most important parameters $T_\infty$ and $\alpha$ can be %assessed 
%estimated from confidence contours in the $\alpha$--$T_\infty$ plane. 
This behavior is illustrated in Figure \ref{fig:T-alpha-contours}, which shows 68.3\%, 95.5\% and 99.7\% confidence contours (for two parameters of interest) for $R_{15}/d_{262}= 1$, 
%\gp{[maybe we can choose $R_\infty =15$ km here?]}
and three values of  color index: $E(B-V)=0.01$, 0.03, and 0.06.
%-- for $R_{15}/d_{262}= 0.87$ 
We see that at $T_4\lesssim 2$--3 the fits become insensitive to the temperature value, and the high-confidence contours fall down to zero temperature, for all the considered color indices.

Figure \ref{fig:T-alpha-contours} also shows that both $T_\infty$ and $\alpha$ grow with increasing $E(B-V)$.
%the best-fit temperature and the spectral index become larger
%goes up, while the best-fit spectral slope becomes more gradual.
The temperature is strongly (anti)correlated with the PL slope, which means that the temperature uncertainty could be reduced significantly if optical (and IR) fluxes were measured more precisely, with additional filters.
% We  see that at 
%larger values of $\alpha$ the fitted temperature becomes so low that the BB component makes no effect on the spectrum. 
 
% This, in turn, means, that if we choose a high confidence level, the temperature is not bounded from below, so that we only can infer an upper limit on $T_\infty$. For instance, at the confidence level of 99\% we obtain $0<T_4 < 10$ and $0<T_4<13$ for 
 %$E(B-V)=0.02$,
% $R_{15}/d_{262}=1$ and 0.8, respectively, at $E(B-V)=0.02$. 
 
 The dependencies of the temperature and spectral slope on color index are shown in Figure \ref{fig:T-alpha-dependencies}. At $E(B-V) < 0.06$ they can be approximated by linear functions given in the figure panels. The fitting parameters also depend on the $R_{15}/d_{262}$ parameter; in particular, the temperature is proportional to $(R_{15}/d_{262})^{-1.5}$  \citep{Pavlov2017}.
 %et al.\ 2017). 
 
Although the probability of very low (undetectable) temperatures is low, such temperatures are not firmly excluded by the current data. Therefore we fit the same data points with a one-component absorbed PL model. For $E(B-V)=0.02$, 
%\old{and $R_{15}/d_{262}$,} 
we obtained 
$\alpha = +0.36\pm 0.06 $, %(\pm0.11)
$f_{\nu_0}=66.7\pm 2.3$ %\, (\pm3.8)$ 
nJy,
%(the uncertainties in parentheses are at the 90\% confidence level) \gp{[do we really need them here? I would remove them.]}
reduced $\chi_{\rm min}^2 = 1.97$ for 4 degrees of freedom. % \yus{showing that this fit is formally worse than that with the BB+PL model}.
%\gp{[one should not compare $\chi^2$ values for different numbers of d.o.f.]}
%\gp{[what are the numbers in parentheses? uncertainties at 90\% level?]} \va{Yes, 68\% (90\%)} 
According to the F-test,  %probability of the PL+BB fit being statistically better 
adding the BB % component 
to the PL component better describes the data  with a probability 
of 0.78. 
%\gp{[correct words? do we really want to write about this?]}
%We also note that none of the pulsars with a measured slope of the IR-optical-UV PL component has so large $\alpha$; in fact, the only pulsar with a positive (but smaller) $\alpha$ is the Crab pulsar, with $\alpha = 0.16\pm0.07$ (Sollerman et al. 2019). This can be considered as an additional argument in favor of the presence of the thermal UV component in the B0950 spectrum.
%\gp{[The last 2 sentences can be moved to Discussion]}

%-- so the best fit is apparently worse than the PL+BB fit. Is the addition of one more spectral parameter justified statistically? Is the F-test applicable for checking?]

%\gp{[how could we evaluate the probability that there is no thermal component at all? Possibly this question is ill-formulated?]}

%Dependence on $R_\infty/d$ ... 
%With account for 
%The combined 
Combining the uncertainties of  extinction and %$R_\infty/d$ 
radius-to-distance ratio %results in a rather broad 
broadens the range of  plausible BB temperatures. For instance, for  
%For plausible $R_\infty/d$ ratio
$0.8 \leqslant R_{15}/d_{262} \leqslant 1.1$ and 
%color excess 
$0.01 \leqslant E(B - V) \leqslant 0.03$, we obtained $5.8 < T_{4} < 12.3$ at the 68\% confidence level, and $4.7 < T_{4} < 13.1$,
%($5.6\times 10^4 < T < 1.8\times 10^5$ K at $M=1.4 M_\odot$) \gp{[could be rounded later]} 
at the 90\% confidence level. 
%\gp{[wouldn't it be reasonable to also provide the range at 68\% conf level?]}
The most conservative temperature upper bound, for 
%$R_{12}/d_{262}= 0.9$ and $E(B - V) = 0.06$ is $T_{4} < 20$ ($T<3.2\times 10^5$ K) 
$R_{15}/d_{262}= 0.8$ and $E(B - V) = 0.06$, is $T_{4} < 17$,
%($T<2.3\times 10^5$ K at $M=1.4 M_\odot$)
at the 99\% confidence level.

%%%%%%%%%%%%%%%%%%%
\section{Discussion \label{disc}}

Thanks to the exquisite angular resolution of the HST, we were able to %exclude the
%contamination of the pulsar image 
resolve 
the pulsar counterpart 
from the close-by red galaxy and 
accurately measure its optical fluxes.
We found that the flux 
in the
%ACS/WFC 
F775W filter
%(SDSS $i$)  = 
(pivot wavelength 7693 \AA) 
is a factor of about 1.5 (1.7) lower than the $I$ ($R$) fluxes previously measured 
%flux densities at close wavelengths, previously measures 
with the VLT with a much worse angular resolution.
The reduced 
flux density in the red part of the spectrum and 
%taking into account the %corrected 
the higher FUV sensitivity inferred from the recent recalibration of the ACS/SBC detector
lead to
%we have obtained 
%the PL + BB fit of the optical-FUV spectrum resulted in 
a more gradual slope of the PL component and a lower temperature of the BB component, in the PL + BB fit of the optical-FUV spectrum.
In particular, the best-fit 
%brightness 
temperature is a factor of 1.5--2 lower than the previous estimates by \citet{Pavlov2017}. For plausible ranges of color index and radius-to-distance ratio, the gravitationally redshifted temperature is in the range $5\lesssim T_\infty/10^4\,{\rm K} \lesssim 13$ (at the 90\% confidence), with a most likely value $T_\infty \sim 8\times 10^4$~K.  The corresponding non-redshifted temperatures are a factor of $1+z = [1-0.246 (M/M_\odot)(12\,{\rm km}/R)]^{-1/2} $ higher; for instance, 
$6 \lesssim T/10^4\,{\rm K} \lesssim 16$ for $M=1.4 M_\odot$ and $R=12$ km. 

%{\bf 
Such temperatures are
%\sout{Being}  
much higher than  
%expected from
predicted by  models of 
passive 
 cooling  of  
 NSs at ages $\ga 10$~Myr.  %late stages of thermal evolution. %for sufficiently old stars. 
 %older than %a few 
 %10 Myrs.
%a 
%\gp{\bf $\sim 17$  Myr old 
%NS (the spin-down age of B0950)}
%\sout{as old as B0950,} 
%\sout{\bf assuming its real age is close to} \sout{\bf the spin-down} \sout{\bf value   
%of 17.5 Myr,}
 If the real age of B0950 is close to its spin-down age of 17.5 Myr,
   %such temperatures 
  this requires heating processes to operate in its  interiors.
  %} 
  According to \cite{GonzalezReisenegger2010}, the most plausible heating mechanism for an ordinary (nonrecycled) pulsar with a strong magnetic field is the so-called vortex creep (or frictional) heating caused by interaction of vortex lines 
of the faster rotating neutron superfluid with the slower rotating normal matter in the inner NS crust. The estimated temperature range is generally consistent with the NS thermal evolution with allowance for the vortex creep heating, for either direct 
%Urca 
or modified Urca cooling 
%(see the lower panels of Figures 4 and 5 in \cite{GonzalezReisenegger2010}. 
(see the lower panels in Figures 4 and 5 in \citet[][]{GonzalezReisenegger2010} 
%{\bf 
and the upper panel in Figure 5  in \citet[][]{2019guillot}).
%}. 

%{\bf
%Another age estimate is the so called kinematic age,  the ratio of the pulsar distance from the Galactic plane, where it is assumed to be born, to its vertical velocity. 
However, the true age of a pulsar can be different from the spin-down age. For instance,
%\sout{As an alternative to the spin-down age,}
using the Bayesian approach, \citet{2019igoshev}  estimated    a most probable  kinematic age of B0950 to be as small as 
%\sout{of  about} 
$\approx 2$ Myr, with a 
%\sout{and its}  
95\% credible interval of 0.4--17 Myr. The upper bound of 
this interval is close to  the spin-down age, requiring the internal heating of the NS, as discussed above. 
Within this very wide interval the agreement 
%\sout{of}
between the derived surface temperature 
%\sout{with}
and the predictions of  passive NS cooling scenarios 
%of the NS 
is only possible for  
ages $\la$ 4 Myr (see Figure 7 in 
\citealt{
2019igoshev}). 
%\sout{It is discussed that}  
%\bf 
According to \citet{2019igoshev}, such small ages %\sout{suggest} 
%\gp{\bf True ages could be  smaller than the spin-down ages} 
could be due to  %\sout{either} 
a fast decay of the NS magnetic field,  
%of the NS  
at a timescale of $\la 5$~Myr,  and/or a  large initial 
spin period of the pulsar, comparable   
with its  current value. %\sout{That appears to make}
Such explanations imply that B0950 is a very peculiar object among other old ordinary pulsars. 
The %\sout{claimed} 
required decay timescale is significantly shorter 
than the Ohmic dissipation  timescale of $\ga 10$ Myr caused by   lattice  impurities of the nuclear crystal in the NS crust 
%(for a recent review of  NS magnetic field evolution 
(see \citealt{2021igoshev} for a recent review of  NS magnetic field evolution). 
Recent 2D and 3D simulations of magneto-thermal evolution of NSs, performed mainly for super-strong magnetic fields $B\sim 10^{14}-10^{15}$~G typical for magnetars, show that short timescales are possible  %\sout{accounting for}
if the non-linear Hall effect 
%\sout{in the magnetic filed evolution equation} 
is taken into account and %\sout{assuming}
the presence of super-strong toroidal field component inside the crust is assumed \citep[e.g.,][]{2014geppert,2018Gourgouliatos}. These  models allow one to explain some observational properties of magnetars. %\sout{While}
However, B0950 does not show any magnetar properties %\sout{like}
such as hard X-ray or soft gamma-ray flares. 
%\sout{or instability of  radio emission, where, however,}  %except of 
%\sout{giant radio pulses as in the Crab-pulsar are observed}
%\sout{at frequencies $\la$ 100 MHz \citep[e.g.,][and reference therein]{2018kazantsev}.} 
%\gp{[I would remove this stuff about radio for simlicity.]}
There are also no signs of the presence of large hot spots on its surface  typical for magnetars. 
%\sout{and incompatible with classical polar hot spots} \sout{ heated by relativistic particles from the pulsar  magnetosphere.} \gp{[again, thses words could be removed for the sake of simplicity]} 
Its surface magnetic field 
%lies in 
is within the range of $\sim 10^{11}-10^{12}$~G, typical for ordinary pulsars. 
The %\sout{suggestion}
assumption of a long initial period, close to the current period of 253 ms, looks rather arbitrary,
%[artificial] %\sout{of B0950 appears to  contradict} 
and it is hardly consistent with much shorter periods of young ordinary pulsars.  
These contradictions do not %\sout{completely} 
exclude the possibility %\sout{that B0950 may be almost an order of magnitude}
of B0950 being younger than
%compared to 
its spin-down age, but 
%\sout{this}
such a large difference in ages appears to be unlikely as %\sout{its physical and radiation} 
the observed properties B0950 are not %\sout{significantly}
substantially different from those of other members of the old pulsar class. 

In addition, %\sout{the estimation of the} 
the small value of the most probable kinematic age of B0950, estimated by \citet{2019igoshev}, is mostly due to the unusually small transverse velocity of this pulsar, which requires a large radial velocity at the assumed probability distribution of total velocities of pulsars. However, we cannot exclude the possibility that the actual total and radial velocities of B0950 are substantially smaller than the most probable values.  
%\sout{is based on assumptions on the pulsar radial} \sout{velocity, which remains unknown. }

%\sout{Nevertheless,}
It would be useful to perform 3D magneto-thermal evolution simulations %\sout{including}
with allowance for the Hall effect for the  field range typical for ordinary pulsars 
%to 
and estimate the timescale of their field decay. As the Hall drift timescale is $\propto B^{-1}$ \citep[e.g.,][]{2021igoshev}, one %\sout{would naively}
can expect a 2-3 orders of magnitude longer drift timescale for %\sout{this case} 
an ordinary pulsar than for a magnetar,
%\sout{as compared to magnetars and the resulting} \sout{time scale for the field decay can be} 
%close/
comparable to the Ohmic 
%one 
field decay time scale, in which case we do not expect a large hot spot on the NS surface to form. 
%\sout{when formation of magnetar like hot region} \sout{might be not effective.}  
On the other hand, simple estimates  \citep[e.g.,][]{GonzalezReisenegger2010} show  that in order to heat B0950 by the magnetic field dissipation to the  temperatures %derived 
inferred from our observations,  its magnetic field 
%has to 
should be   $\sim 10^{13}$~G, much higher than the spin-down value of $ 2.4\times 10^{11}$~G.     
%}  

To estimate the parameters that determine the rate of the vortex creep heating (such as the so-called ``excess angular momentum'' 
 -- see, e.g., \citealt{Gugercinoglu2017} for a recent discussion), more pulsars with ages $\gtrsim$ a few Myr should be observed in the FUV-optical range to estimate the NS temperatures and understand the thermal evolution of old NSs. Until new space observatories with UV capabilities are launched, such observation are only possible with the HST. 
%\old{Unfortunately, ....} 

%are comparable to (slightly lower than) the BB temperature of the millisecond pulsar J0437--4715 %(Kargaltsev et al.\ 2004; 
%\citep{Kargaltsev2004, Durant2012, Guillot2016}, the first recycled pulsar whose 

The spectral 
slope of the PL component found from the HST observations, $\alpha = -0.3\pm0.3$, is not nearly as steep as
$\alpha=-1.2\pm0.4$ based on the previous ground-based observations. %$\alpha= %-0.32^{+0.29}_{-0.26}$ 
%-0.3\pm0.3$ versus 
%$\alpha=-1.17\pm0.43$. 
%\alpha=-1.2\pm 0.4$. 
%Such a 
The newly measured slope is within the range of 
%\old{$-0.7[??] \lesssim \alpha \lesssim 0.1$,} 
%\yus{
%\sout{$-0.7 \lesssim \alpha \lesssim 0.2$,} 
$-1 \lesssim \alpha \lesssim 0$, 
%\gp{[maybe $-1\lesssim \alpha\lesssim 0$ ?]}
found for the optical  spectra of other rotation-powered pulsars (e.g., \citealt{mignani2007}). 
%\sout{\citet{Mignani2019}).} 
%\gp{[I see no the former limits in the cited papers by Mignani. In Mignani et al (2019) it is written 
%$-1\lesssim \alpha \lesssim 0$, with a ref to Mignani 2011. 
%I am not sure what should we do... Maybe indeed 
%I suggest to change to $-1\lesssim \alpha \lesssim 0$ and refer to Mignani 2011?]} 
%\yus{Seems, the most  appropriate ref is \citep{mignani2007}, where  optical PL indices are really collected and partially analysed. }
The optical PL slope of B0950 is flatter 
%\old{(more gradual)} 
than in the X-ray range -- e.g., $\alpha_X=-0.75\pm 0.15$ for a single-component PL fit  \citep{ZavlinPavlov04}, which implies a spectral break between the optical and X-rays, similar to other  pulsars investigated in both ranges.
%
%{\bf 
However, the break disappears if the thermal component from 
pulsar hot polar caps heated by relativistic  particles from the pulsar magnetosphere contributes to X-ray emission, leading to  $\alpha_X=-0.3\pm 0.1$, which is consistent with the optical slope.
It is thus unclear whether or not the optical and
X-ray nonthermal photons are  produced by the same emission mechanism 
%(synchrotron, curvature, synchrocurvture} 
and the same relativistic particle 
distribution in the pulsar magnetosphere.
%, or not.}
%\sout{A model of synchro-curvature radiation of} \sout{ particles accelerated in pulsar magnetosphere} 
%\sout{appears to be able to fit  phase-averaged spectra}
%\sout{of multiwavelength PL emission   of  pulsars}  \sout{from optical to $\gamma$-rays \citep[e.g.,][]{Torres2019}.}
%\gp{[see my remarks in the response.]} \sout{According to the model,  synchrotron emission}  \sout{dominates at long wavelengths, while the curvature} \sout{radiation dominates at higher photon energies.} 
%\gp{[It had been clear long ago, before this model appeared.]} 
%\sout{Large uncertainties of the PL slopes in} \sout{the optical-UV and X-rays makes useless invoking}  \sout{this  model 
%does not allow one to get any meaningful 
% constraints on the parameters of the synchro-curvature model 
% for B0950 at current data stage. It} \sout{may be tried later using new  deep X-ray observations} \sout{of B0950 performed by us recently with XMM-Newton,} \sout{which can help to firmly establish the presence} \sout{of the hot caps component and at least  decrease} \sout{uncertainties of the PL parameters in X-rays.}   
% \gp{[No need to describe here (inapplicable) models. Adding X-rays would not be enough to use that model -- it is not a gamma-ray pulsar. I see no need to inform the reader about our new XMM observations, some of which are public now, and the others will become public soon.]}}
  The properties of the polar caps and their contribution to the X-ray spectrum can be inferred from timing and spectral analysis of deeper X-ray observations.
  %}

Although 
%the presence of the thermal component with $T_4\sim 10$ 
it is quite plausible that the FUV flux of B0950 is dominated by the thermal component,
%with $T_4\sim 10$ in the optical-FUV spectrum of B0950 looks very plausible, 
we cannot firmly exclude the possibility of the temperature being lower, and the spectral index $\alpha$ larger, than estimated above 
%and the contribution of the thermal component is negligible.  %contribution is so small spectral slope $\alpha$
(see Figure  \ref{fig:T-alpha-contours}). In other words, if we choose a high confidence level for the fit parameter uncertainties, we can only  put an upper limit on the brightness temperature. For instance,
at the confidence level of 99\%, we obtained $T_\infty < 1.0\times 10^5$ K at $E(B-V)=0.02$ and $R_{15}/d_{262}=1$, and $T_\infty < 1.7\times 10^5$ K at $E(B-V) = 0.06$ and $R_{15}/d_{262}=0.8$. Given the dependencies of $T_\infty$ on $E(B-V)$ and $R_\infty/d$ (see Section \ref{sec:spectralfits}), the value $T_\infty = 1.7\times 10^5$ K 
%(or $T= 2.1\times 10^5$ K at $M=1.4 M_\odot$ and $R=12$ km \va{$T= 2.1\times 10^5$ K at $M=1.1 M_\odot$ and $R= 10$ km ?}) 
can be considered as the most conservative upper limit on the temperature.

%: $\alpha=-0.32^{+0.29}_{-0.26}$ versus $\alpha = -1.17\pm 0.43$, and a lower brightness temperature $T_4 = 8.0^{+0.9}_{-1.4}$ ...

Because of the $\alpha$-$T_\infty$ anti-correlation (see Figure \ref{fig:T-alpha-contours}), very low temperature values correspond to positive values of $\alpha$, i.e., the flux density increasing with frequency (e.g., $\alpha = +0.36\pm0.06$ at $E(B-V)=0.02$).
%\yus{and single PL model}). 
Not only the low-temperature solutions are less acceptable statistically, but also none of the pulsars with a measured slope of the optical PL component shows so large $\alpha$; the only pulsar with a positive (but smaller) $\alpha$ is the Crab pulsar, with $\alpha=0.16\pm0.07$ \citep{Sollerman19}. This can be considered as an additional argument in favor of the presence of the thermal UV component in the B0950 spectrum.

To conclude, 
%the analysis of 
our observations of B0950 with the HST %have
%shown that the 
%confirmed the likely 
support the presence of the thermal UV component, 
%with temperatures around $1\times 10^5$ K,
whose temperature, however, is lower than previously suggested. They have also shown that the slope of the nonthermal spectral component, presumably emitted from the pulsar  magnetosphere, is not as steep as found from groundbased  observations. To firmly prove the presence of the hot thermal emission and measure the temperature and PL slope more precisely, deep HST and/or JWST observation in several IR-optical-UV filters
%with the HST and/or JWST 
are needed. To study the thermal and magnetospheric evolution of NSs, it
would be very important to observe a large sample of isolated NSs of various types, employing the unique UV capabilities of the HST.
%for the study of a larger sample of isolated NSs of various types.

\begin{acknowledgments}
%{\bf 
We thank the anonymous referee for the useful remarks.
%}
Support for program \#16064 was provided by NASA through a grant from the Space Telescope Science Institute, which is operated by the Association of Universities for Research in Astronomy, Inc., under NASA contract NAS 5-26555.
\end{acknowledgments}

\facilities{HST(ACS/WFC), HST(WFC3/UVIS)}

%{\bf 
Some of the data presented in this paper were obtained from the Mikulski Archive for Space Telescopes (MAST) at the Space Telescope Science Institute. The specific observations analyzed can be accessed via \dataset[10.17909/t9-e9a9-4c40]{https://doi.org/10.17909/t9-e9a9-4c40}.
%}

\software{DrizzlePac %(\url{http://drizzlepac.stsci.edu/})
%{\bf 
\citep{Hack2013}
%}, 
IRAF \citep{1986Tody, Tody1993}}\\

\bibliographystyle{aasjournal}
\bibliography{B0950}

\begin{thebibliography}{}
\expandafter\ifx\csname natexlab\endcsname\relax\def\natexlab#1{#1}\fi
\providecommand{\url}[1]{\href{#1}{#1}}
\providecommand{\dodoi}[1]{doi:~\href{http://doi.org/#1}{\nolinkurl{#1}}}
\providecommand{\doeprint}[1]{\href{http://ascl.net/#1}{\nolinkurl{http://ascl.net/#1}}}
\providecommand{\doarXiv}[1]{\href{https://arxiv.org/abs/#1}{\nolinkurl{https://arxiv.org/abs/#1}}}

\bibitem[{{Alpar} {et~al.}(1984){Alpar}, {Pines}, {Anderson}, \&
  {Shaham}}]{Alpar1984}
{Alpar}, M.~A., {Pines}, D., {Anderson}, P.~W., \& {Shaham}, J. 1984, \apj,
  276, 325, \dodoi{10.1086/161616}

\bibitem[{{Avila} {et~al.}(2019){Avila}, {Bohlin}, {Hathi}, {Lockwood}, {Lim},
  \& {De La Pena}}]{Avila2019}
{Avila}, R.~J., {Bohlin}, R., {Hathi}, N., {et~al.} 2019, {SBC Absolute Flux
  Calibration}, Tech. rep.

\bibitem[{{Brisken} {et~al.}(2002){Brisken}, {Benson}, {Goss}, \&
  {Thorsett}}]{Brisken2002}
{Brisken}, W.~F., {Benson}, J.~M., {Goss}, W.~M., \& {Thorsett}, S.~E. 2002,
  \apj, 571, 906, \dodoi{10.1086/340098}

\bibitem[{{Cardelli} {et~al.}(1989){Cardelli}, {Clayton}, \&
  {Mathis}}]{Cardelli1989}
{Cardelli}, J.~A., {Clayton}, G.~C., \& {Mathis}, J.~S. 1989, \apj, 345, 245,
  \dodoi{10.1086/167900}

\bibitem[{{Casertano} {et~al.}(2000){Casertano}, {de Mello}, {Dickinson},
  {Ferguson}, {Fruchter}, {Gonzalez-Lopezlira}, {Heyer}, {Hook}, {Levay},
  {Lucas}, {Mack}, {Makidon}, {Mutchler}, {Smith}, {Stiavelli}, {Wiggs}, \&
  {Williams}}]{Casertano2000}
{Casertano}, S., {de Mello}, D., {Dickinson}, M., {et~al.} 2000, \aj, 120,
  2747, \dodoi{10.1086/316851}

\bibitem[{{Deustua} {et~al.}(2017){Deustua}, {Mack}, {Bajaj}, \&
  {Khandrika}}]{Deustua2017}
{Deustua}, S.~E., {Mack}, J., {Bajaj}, V., \& {Khandrika}, H. 2017, {WFC3/UVIS
  Updated 2017 Chip-Dependent Inverse Sensitivity Values}, Space Telescope WFC
  Instrument Science Report

\bibitem[{{Gaia Collaboration} {et~al.}(2021){Gaia Collaboration}, {Brown},
  {Vallenari}, {Prusti}, {de Bruijne}, {Babusiaux}, {Biermann}, {Creevey},
  {Evans}, {Eyer}, {Hutton}, {Jansen}, {Jordi}, {Klioner}, {Lammers},
  {Lindegren}, {Luri}, {Mignard}, {Panem}, {Pourbaix}, {Randich}, {Sartoretti},
  {Soubiran}, {Walton}, {Arenou}, {Bailer-Jones}, {Bastian}, {Cropper},
  {Drimmel}, {Katz}, {Lattanzi}, {van Leeuwen}, {Bakker}, {Cacciari},
  {Casta{\~n}eda}, {De Angeli}, {Ducourant}, {Fabricius}, {Fouesneau},
  {Fr{\'e}mat}, {Guerra}, {Guerrier}, {Guiraud}, {Jean-Antoine Piccolo},
  {Masana}, {Messineo}, {Mowlavi}, {Nicolas}, {Nienartowicz}, {Pailler},
  {Panuzzo}, {Riclet}, {Roux}, {Seabroke}, {Sordo}, {Tanga}, {Th{\'e}venin},
  {Gracia-Abril}, {Portell}, {Teyssier}, {Altmann}, {Andrae}, {Bellas-Velidis},
  {Benson}, {Berthier}, {Blomme}, {Brugaletta}, {Burgess}, {Busso}, {Carry},
  {Cellino}, {Cheek}, {Clementini}, {Damerdji}, {Davidson}, {Delchambre},
  {Dell'Oro}, {Fern{\'a}ndez-Hern{\'a}ndez}, {Galluccio}, {Garc{\'\i}a-Lario},
  {Garcia-Reinaldos}, {Gonz{\'a}lez-N{\'u}{\~n}ez}, {Gosset}, {Haigron},
  {Halbwachs}, {Hambly}, {Harrison}, {Hatzidimitriou}, {Heiter},
  {Hern{\'a}ndez}, {Hestroffer}, {Hodgkin}, {Holl}, {Jan{\ss}en}, {Jevardat de
  Fombelle}, {Jordan}, {Krone-Martins}, {Lanzafame}, {L{\"o}ffler}, {Lorca},
  {Manteiga}, {Marchal}, {Marrese}, {Moitinho}, {Mora}, {Muinonen}, {Osborne},
  {Pancino}, {Pauwels}, {Petit}, {Recio-Blanco}, {Richards}, {Riello},
  {Rimoldini}, {Robin}, {Roegiers}, {Rybizki}, {Sarro}, {Siopis}, {Smith},
  {Sozzetti}, {Ulla}, {Utrilla}, {van Leeuwen}, {van Reeven}, {Abbas}, {Abreu
  Aramburu}, {Accart}, {Aerts}, {Aguado}, {Ajaj}, {Altavilla}, {{\'A}lvarez},
  {{\'A}lvarez Cid-Fuentes}, {Alves}, {Anderson}, {Anglada Varela}, {Antoja},
  {Audard}, {Baines}, {Baker}, {Balaguer-N{\'u}{\~n}ez}, {Balbinot}, {Balog},
  {Barache}, {Barbato}, {Barros}, {Barstow}, {Bartolom{\'e}}, {Bassilana},
  {Bauchet}, {Baudesson-Stella}, {Becciani}, {Bellazzini}, {Bernet}, {Bertone},
  {Bianchi}, {Blanco-Cuaresma}, {Boch}, {Bombrun}, {Bossini}, {Bouquillon},
  {Bragaglia}, {Bramante}, {Breedt}, {Bressan}, {Brouillet}, {Bucciarelli},
  {Burlacu}, {Busonero}, {Butkevich}, {Buzzi}, {Caffau}, {Cancelliere},
  {C{\'a}novas}, {Cantat-Gaudin}, {Carballo}, {Carlucci}, {Carnerero},
  {Carrasco}, {Casamiquela}, {Castellani}, {Castro-Ginard}, {Castro Sampol},
  {Chaoul}, {Charlot}, {Chemin}, {Chiavassa}, {Cioni}, {Comoretto}, {Cooper},
  {Cornez}, {Cowell}, {Crifo}, {Crosta}, {Crowley}, {Dafonte}, {Dapergolas},
  {David}, {David}, {de Laverny}, {De Luise}, {De March}, {De Ridder}, {de
  Souza}, {de Teodoro}, {de Torres}, {del Peloso}, {del Pozo}, {Delbo},
  {Delgado}, {Delgado}, {Delisle}, {Di Matteo}, {Diakite}, {Diener},
  {Distefano}, {Dolding}, {Eappachen}, {Edvardsson}, {Enke}, {Esquej}, {Fabre},
  {Fabrizio}, {Faigler}, {Fedorets}, {Fernique}, {Fienga}, {Figueras},
  {Fouron}, {Fragkoudi}, {Fraile}, {Franke}, {Gai}, {Garabato},
  {Garcia-Gutierrez}, {Garc{\'\i}a-Torres}, {Garofalo}, {Gavras}, {Gerlach},
  {Geyer}, {Giacobbe}, {Gilmore}, {Girona}, {Giuffrida}, {Gomel}, {Gomez},
  {Gonzalez-Santamaria}, {Gonz{\'a}lez-Vidal}, {Granvik},
  {Guti{\'e}rrez-S{\'a}nchez}, {Guy}, {Hauser}, {Haywood}, {Helmi}, {Hidalgo},
  {Hilger}, {H{\l}adczuk}, {Hobbs}, {Holland}, {Huckle}, {Jasniewicz},
  {Jonker}, {Juaristi Campillo}, {Julbe}, {Karbevska}, {Kervella}, {Khanna},
  {Kochoska}, {Kontizas}, {Kordopatis}, {Korn}, {Kostrzewa-Rutkowska},
  {Kruszy{\'n}ska}, {Lambert}, {Lanza}, {Lasne}, {Le Campion}, {Le Fustec},
  {Lebreton}, {Lebzelter}, {Leccia}, {Leclerc}, {Lecoeur-Taibi}, {Liao},
  {Licata}, {Lindstr{\o}m}, {Lister}, {Livanou}, {Lobel}, {Madrero Pardo},
  {Managau}, {Mann}, {Marchant}, {Marconi}, {Marcos Santos}, {Marinoni},
  {Marocco}, {Marshall}, {Martin Polo}, {Mart{\'\i}n-Fleitas}, {Masip},
  {Massari}, {Mastrobuono-Battisti}, {Mazeh}, {McMillan}, {Messina},
  {Michalik}, {Millar}, {Mints}, {Molina}, {Molinaro}, {Moln{\'a}r},
  {Montegriffo}, {Mor}, {Morbidelli}, {Morel}, {Morris}, {Mulone}, {Munoz},
  {Muraveva}, {Murphy}, {Musella}, {Noval}, {Ord{\'e}novic}, {Orr{\`u}},
  {Osinde}, {Pagani}, {Pagano}, {Palaversa}, {Palicio}, {Panahi}, {Pawlak},
  {Pe{\~n}alosa Esteller}, {Penttil{\"a}}, {Piersimoni}, {Pineau}, {Plachy},
  {Plum}, {Poggio}, {Poretti}, {Poujoulet}, {Pr{\v{s}}a}, {Pulone}, {Racero},
  {Ragaini}, {Rainer}, {Raiteri}, {Rambaux}, {Ramos}, {Ramos-Lerate}, {Re
  Fiorentin}, {Regibo}, {Reyl{\'e}}, {Ripepi}, {Riva}, {Rixon}, {Robichon},
  {Robin}, {Roelens}, {Rohrbasser}, {Romero-G{\'o}mez}, {Rowell}, {Royer},
  {Rybicki}, {Sadowski}, {Sagrist{\`a} Sell{\'e}s}, {Sahlmann}, {Salgado},
  {Salguero}, {Samaras}, {Sanchez Gimenez}, {Sanna}, {Santove{\~n}a},
  {Sarasso}, {Schultheis}, {Sciacca}, {Segol}, {Segovia}, {S{\'e}gransan},
  {Semeux}, {Shahaf}, {Siddiqui}, {Siebert}, {Siltala}, {Slezak}, {Smart},
  {Solano}, {Solitro}, {Souami}, {Souchay}, {Spagna}, {Spoto}, {Steele},
  {Steidelm{\"u}ller}, {Stephenson}, {S{\"u}veges}, {Szabados}, {Szegedi-Elek},
  {Taris}, {Tauran}, {Taylor}, {Teixeira}, {Thuillot}, {Tonello}, {Torra},
  {Torra}, {Turon}, {Unger}, {Vaillant}, {van Dillen}, {Vanel}, {Vecchiato},
  {Viala}, {Vicente}, {Voutsinas}, {Weiler}, {Wevers}, {Wyrzykowski}, {Yoldas},
  {Yvard}, {Zhao}, {Zorec}, {Zucker}, {Zurbach}, \&
  {Zwitter}}]{GaiaCollaboration2021}
{Gaia Collaboration}, {Brown}, A.~G.~A., {Vallenari}, A., {et~al.} 2021, \aap,
  649, A1, \dodoi{10.1051/0004-6361/202039657}

\bibitem[{{Geppert} \& {Vigan{\`o}}(2014)}]{2014geppert}
{Geppert}, U., \& {Vigan{\`o}}, D. 2014, \mnras, 444, 3198,
  \dodoi{10.1093/mnras/stu1675}

\bibitem[{{Gonzalez} \& {Reisenegger}(2010)}]{GonzalezReisenegger2010}
{Gonzalez}, D., \& {Reisenegger}, A. 2010, \aap, 522, A16,
  \dodoi{10.1051/0004-6361/201015084}

\bibitem[{{Gourgouliatos} \& {Hollerbach}(2018)}]{2018Gourgouliatos}
{Gourgouliatos}, K.~N., \& {Hollerbach}, R. 2018, \apj, 852, 21,
  \dodoi{10.3847/1538-4357/aa9d93}

\bibitem[{{G{\"u}gercino{\u{g}}lu}(2017)}]{Gugercinoglu2017}
{G{\"u}gercino{\u{g}}lu}, E. 2017, \mnras, 469, 2313,
  \dodoi{10.1093/mnras/stx985}

\bibitem[{{Guillot} {et~al.}(2019){Guillot}, {Pavlov}, {Reyes}, {Reisenegger},
  {Rodriguez}, {Rangelov}, \& {Kargaltsev}}]{2019guillot}
{Guillot}, S., {Pavlov}, G.~G., {Reyes}, C., {et~al.} 2019, \apj, 874, 175,
  \dodoi{10.3847/1538-4357/ab0f38}

\bibitem[{{Hack} {et~al.}(2013){Hack}, {Dencheva}, \& {Fruchter}}]{Hack2013}
{Hack}, W.~J., {Dencheva}, N., \& {Fruchter}, A.~S. 2013, in Astronomical
  Society of the Pacific Conference Series, Vol. 475, Astronomical Data
  Analysis Software and Systems XXII, ed. D.~N. {Friedel}, 49

\bibitem[{{Igoshev}(2019)}]{2019igoshev}
{Igoshev}, A.~P. 2019, \mnras, 482, 3415, \dodoi{10.1093/mnras/sty2945}

\bibitem[{{Igoshev} {et~al.}(2021){Igoshev}, {Popov}, \&
  {Hollerbach}}]{2021igoshev}
{Igoshev}, A.~P., {Popov}, S.~B., \& {Hollerbach}, R. 2021, Universe, 7, 351,
  \dodoi{10.3390/universe7090351}

\bibitem[{{Larson} \& {Link}(1999)}]{LarsonLink1999}
{Larson}, M.~B., \& {Link}, B. 1999, \apj, 521, 271, \dodoi{10.1086/307532}

\bibitem[{{Manchester} {et~al.}(2005){Manchester}, {Hobbs}, {Teoh}, \&
  {Hobbs}}]{Manchester2005}
{Manchester}, R.~N., {Hobbs}, G.~B., {Teoh}, A., \& {Hobbs}, M. 2005, VizieR
  Online Data Catalog, VII/245

\bibitem[{{Merline} \& {Howell}(1995)}]{MerlineHowell1995}
{Merline}, W.~J., \& {Howell}, S.~B. 1995, Experimental Astronomy, 6, 163,
  \dodoi{10.1007/BF00421131}

\bibitem[{{Mignani} {et~al.}(2007){Mignani}, {Zharikov}, \&
  {Caraveo}}]{mignani2007}
{Mignani}, R.~P., {Zharikov}, S., \& {Caraveo}, P.~A. 2007, \aap, 473, 891,
  \dodoi{10.1051/0004-6361:20077774}

\bibitem[{{Page} {et~al.}(2009){Page}, {Lattimer}, {Prakash}, \&
  {Steiner}}]{Page2009}
{Page}, D., {Lattimer}, J.~M., {Prakash}, M., \& {Steiner}, A.~W. 2009, \apj,
  707, 1131, \dodoi{10.1088/0004-637X/707/2/1131}

\bibitem[{{Pavlov}(1992)}]{Pavlov1992}
{Pavlov}, G.~G. 1992, in IAU Colloq. 128: Magnetospheric Structure and Emission
  Mechanics of Radio Pulsars, ed. T.~H. {Hankins}, J.~M. {Rankin}, \& J.~A.
  {Gil}, 220

\bibitem[{{Pavlov} {et~al.}(2017){Pavlov}, {Rangelov}, {Kargaltsev},
  {Reisenegger}, {Guillot}, \& {Reyes}}]{Pavlov2017}
{Pavlov}, G.~G., {Rangelov}, B., {Kargaltsev}, O., {et~al.} 2017, \apj, 850,
  79, \dodoi{10.3847/1538-4357/aa947c}

\bibitem[{{Pavlov} {et~al.}(1996){Pavlov}, {Stringfellow}, \&
  {Cordova}}]{Pavlov1996}
{Pavlov}, G.~G., {Stringfellow}, G.~S., \& {Cordova}, F.~A. 1996, \apj, 467,
  370, \dodoi{10.1086/177612}

\bibitem[{{Pilkington} {et~al.}(1968){Pilkington}, {Hewish}, {Bell}, \&
  {Cole}}]{1968Natur}
{Pilkington}, J.~D.~H., {Hewish}, A., {Bell}, S.~J., \& {Cole}, T.~W. 1968,
  \nat, 218, 126, \dodoi{10.1038/218126a0}

\bibitem[{{Shibazaki} \& {Lamb}(1989)}]{ShibazakiLamb1989}
{Shibazaki}, N., \& {Lamb}, F.~K. 1989, \apj, 346, 808, \dodoi{10.1086/168062}

\bibitem[{{Skelton} {et~al.}(2014){Skelton}, {Whitaker}, {Momcheva}, {Brammer},
  {van Dokkum}, {Labb{\'e}}, {Franx}, {van der Wel}, {Bezanson}, {Da Cunha},
  {Fumagalli}, {F{\"o}rster Schreiber}, {Kriek}, {Leja}, {Lundgren}, {Magee},
  {Marchesini}, {Maseda}, {Nelson}, {Oesch}, {Pacifici}, {Patel}, {Price},
  {Rix}, {Tal}, {Wake}, \& {Wuyts}}]{Skelton2014}
{Skelton}, R.~E., {Whitaker}, K.~E., {Momcheva}, I.~G., {et~al.} 2014, \apjs,
  214, 24, \dodoi{10.1088/0067-0049/214/2/24}

\bibitem[{{Sollerman} {et~al.}(2019){Sollerman}, {Selsing}, {Vreeswijk},
  {Lundqvist}, \& {Nyholm}}]{Sollerman19}
{Sollerman}, J., {Selsing}, J., {Vreeswijk}, P.~M., {Lundqvist}, P., \&
  {Nyholm}, A. 2019, \aap, 629, A140, \dodoi{10.1051/0004-6361/201935086}

\bibitem[{{Tody}(1986)}]{1986Tody}
{Tody}, D. 1986, in Society of Photo-Optical Instrumentation Engineers (SPIE)
  Conference Series, Vol. 627, Instrumentation in astronomy VI, ed. D.~L.
  {Crawford}, 733, \dodoi{10.1117/12.968154}

\bibitem[{{Tody}(1993)}]{Tody1993}
{Tody}, D. 1993, in Astronomical Society of the Pacific Conference Series,
  Vol.~52, Astronomical Data Analysis Software and Systems II, ed. R.~J.
  {Hanisch}, R.~J.~V. {Brissenden}, \& J.~{Barnes}, 173

\bibitem[{{Yakovlev} \& {Pethick}(2004)}]{YakovlevPethick2004}
{Yakovlev}, D.~G., \& {Pethick}, C.~J. 2004, \araa, 42, 169,
  \dodoi{10.1146/annurev.astro.42.053102.134013}

\bibitem[{{Zavlin} \& {Pavlov}(2004)}]{ZavlinPavlov04}
{Zavlin}, V.~E., \& {Pavlov}, G.~G. 2004, \apj, 616, 452,
  \dodoi{10.1086/424894}

\bibitem[{{Zharikov} {et~al.}(2004){Zharikov}, {Shibanov}, {Mennickent},
  {Komarova}, {Koptsevich}, \& {Tovmassian}}]{Zharikov2004}
{Zharikov}, S.~V., {Shibanov}, Y.~A., {Mennickent}, R.~E., {et~al.} 2004, \aap,
  417, 1017, \dodoi{10.1051/0004-6361:20034230}

\bibitem[{{Zharikov} {et~al.}(2002){Zharikov}, {Shibanov}, {Koptsevich},
  {Kawai}, {Urata}, {Komarova}, {Sokolov}, {Shibata}, \&
  {Shibazaki}}]{Zharikov2002}
{Zharikov}, S.~V., {Shibanov}, Y.~A., {Koptsevich}, A.~B., {et~al.} 2002, \aap,
  394, 633, \dodoi{10.1051/0004-6361:20021155}

\end{thebibliography}

\end{document}